\documentclass[fleqn,usenatbib]{mnras}
\usepackage{newtxtext,newtxmath}
\usepackage[T1]{fontenc}

\DeclareRobustCommand{\VAN}[3]{#2}
\let\VANthebibliography\thebibliography
\def\thebibliography{\DeclareRobustCommand{\VAN}[3]{##3}\VANthebibliography}

\usepackage{graphicx}	% Including figure files
\usepackage{amsmath}	% Advanced maths commands
\usepackage{verbatim}   % To use begin{comment} for removing blocks of text
\usepackage{xcolor}

\title[Extended Corona Models of X-ray Reverberation in AGN]{Extended Corona Models of X-ray Reverberation in the AGN 1H~0707-495 and IRAS 13224-3809}

\author[S. Hancock et al.]{
S. Hancock,$^{1}$\thanks{E-mail: steff.hancock@bristol.ac.uk}
A. J. Young,$^{1}$
P. Chainakun$^{2,3}$
\\
$^{1}$HH Wills Physics Laboratory, Tyndall Avenue, Bristol BS8 1TL, UK\\
$^{2}$School of Physics, Institute of Science, Suranaree University of Technology, Nakhon Ratchasima 30000, Thailand\\
$^3$Centre of Excellence in High Energy Physics and Astrophysics, Suranaree University of Technology, Nakhon Ratchasima 30000, Thailand\\
}

\date{Accepted XXX. Received YYY; in original form ZZZ}

\pubyear{2020}

\begin{document}
\label{firstpage}
\pagerange{\pageref{firstpage}--\pageref{lastpage}}
\maketitle

\begin{abstract}
We fit a new vertically extended corona model to previously measured reverberation time lags observed by \emph{XMM-Newton} in two extremely variable Narrow Line Seyfert 1 Active Galactic Nuclei (AGN), 1H~0707-495 and IRAS~13224-3809, in a variety of similarly observed flux groups and explore the model in all observations over a 16 year period. The model employs two X-ray sources located along the black hole rotational axis at height, $h_1$ and $h_2$ respectively. These sources have their associated photon indices $\Gamma_1$ and $\Gamma_2$ which respond to fluctuations in the disc with a maximum response duration of $t_\text{max}$ and a propagation delay between the response of the two of $t_\text{shift}$. We find that for 1H 0707-495, $h_2$ is significantly correlated with $\Gamma_1$ and anti-correlated with ionisation $\xi$. Whilst the 1H 0707-495 corona extends upwards, the emission appears softer and the disc is less ionised. We find similarities in IRAS 13224-3809, but significant anti-correlation between $\Gamma_2$ and both $t_\text{max}$ and $t_\text{shift}$. This suggests that when the IRAS 13224-3809 corona becomes softer while extending vertically upwards, the overall corona response occurs faster. This may also suggest that the inner disc also becomes more active. In addition, $\Gamma_1$ and $\Gamma_2$ are extreme, relatively less variable, but more separate in IRAS 13224-3809 than in 1H 0707-495. This suggests that the IRAS 13224-3809 corona may be more patchy in the sense that it has two more clear distinct spectral zones of $\Gamma_1$ and $\Gamma_2$ (possibly relating to two distinct zones of coronal temperature) when compared to 1H 0707-495.

\end{abstract}


\section{Introduction}
The spectra of active galactic nuclei (AGN) have a component of direct X-ray continuum emission, and that cold gas can reflect some of this X-ray continuum \citep{1990Natur.344..132P}. The cold, optically thick material seen through fluorescence and reflection is known to occur in the presence of an accretion disc where X-rays produced by inverse Compton scattering in a corona are reflected off the accretion disc producing a modified spectrum emission with fluorescent Fe K lines at 6.4\,keV and other spectral features \citep{1991MNRAS.249..352G}. The first hints of reflection or reprocessing time-delay due to the light travel time between the corona and disk were seen in \emph{XMM-Newton} observations of Ark564 \citep{2007MNRAS.382..985M} which led to the first robustly discovered delays in 1H0707-495 \citep{2009Natur.459..540F} where the soft energy band ($0.3-1$\,keV) lagged behind the hard band ($1-4$\,keV) by 30\,s. Many reverberation lags have since been discovered \citep[e.g.][]{2011MNRAS.416L..94E,2013MNRAS.434.1129K,2013ApJ...767..121Z}. 

The soft negative time lag is the signature of relativistic reflection that reverberates in response to continuum fluctuations and is interpreted as the light crossing time from the source to the reflecting region, correlating positively with the black hole mass \citep{2013MNRAS.431.2441D}. Hard positive lags at lower frequencies are understood to originate from fluctuations of the accretion rate propagating from outer to inner radii, causing the hard X-rays produced at smaller radii to respond after soft X-rays produced at larger radii \cite[e.g.][]{2001MNRAS.327..799K,2006MNRAS.367..801A,2008MNRAS.388..211A}. Hard X-ray lags (i.e. hard photon variability lagging soft photon variability) were evident in X-ray binaries before they were discovered in AGN \cite[see e.g.][]{1988Natur.336..450M,Nowak1999}.

The reproduction of soft reverberation lags from a compact corona and the hard lags produced by propagating fluctuations through an extended region whilst maintaining the features of the energy dependence seen in the Fe K$\alpha$ line region is challenging and computationally intensive. Motivated by these phenomena, \cite{chainakun_investigating_2017}, CY17 hereafter, developed an approximation of a vertically extended source using two X-ray point sources located on the rotation axis of the black hole. The model employs gravitational units where the gravitational radius $r_g = GM/c^2$ and gravitational time $t_g = GM/c^3$ (where $G$ is the gravitational constant, $M$ is the black hole mass and $c$ is the speed of light). Instead of modelling the propagating fluctuations, a phenomenological function of expected source responses which react to these propagations is employed. The two X-ray sources are allowed to vary as they respond to primary intrinsic variations. The X-ray continuum variability depends on the  response of the source and the X-ray reflection depends also on the disc response. Therefore the model has to predict the time lags from both continuum X-ray sources and the associated disc responses. The model was fitted to a 120 ks timing \textit{XMM-Newton} observation of the narrow-line Seyfert 1 galaxy PG~1244+026, examining the frequency and energy where the lags were found. The model revealed hard and soft X-ray sources at heights $\sim6\ r_g$ and $\sim11 \ r_g$ respectively with the upper source producing small amounts of reflection which suggested a feasible geometry of a relativistic jet beaming away from the disc. CY17 suggested that the continuum flux from the upper source and the extra blackbody component that contribute significant flux at energies $< 1$\,keV are required to dilute the soft reverberation lags and to reproduce the absence of soft lag in the lag-energy spectrum of PG~1244+026.

The X-ray variability in 1H~0707-495 was found to be extreme and may have both intrinsic and environmental absorption origins \citep{2021MNRAS.508.1798P}. The geometry of 1H~0707-495 has been discussed by \cite{2020A&A...641A..89S} who developed a new extended lamppost model which accounted for the spatial extent and rotation of the X-ray source. The investigation of the location and size of the corona indicated a compact corona at $\sim 2\ r_g$ (highly centrally peaked rather than extended) regardless of the effects of ionised absorption from winds. They found no evidence that the size of the corona was correlated to the luminosity as reported by \cite{2014MNRAS.443.2746W}. The time lags in IRAS~13224-3809 have been discussed by \cite{alston_dynamic_2020} who studied short-timescale variations and included all relativistic effects allowing for ultra-fast outflows fitting multiple epochs where the source height changed. They tackled inherent degeneracies between the reverberation signal and black hole mass, inner disc radius and height of the corona by tracking short-scale reverberation signatures where the source height changed. They found that the height of the corona increased with increasing luminosity and that black hole mass uncertainty estimates were comparable to the leading optical reverberation method by \cite{Peterson_2004}. 
In addition, \cite{2020MNRAS.498.3184C} used all available data from the \textit{XMM-Netwon} archive to simultaneously fit various flux states of the energy spectra and time lags using a new code which calculates reflection spectra from the accretion disc in response to an X–ray flare from a point source located above the black hole in accretion disc lamp-post geometry. The model strongly favoured a maximally spinning black hole and detected significant variations of the corona height, increasing from $3 - 5\ r_g$ at lower flux states and extending to $\sim10 - 20\ r_g$ when the luminosity doubled. Recently, \cite{2022ApJ...934..166C} constrained the reverberation signatures that appeared in the power spectral density of IRAS~13224-3809 and found that the lamp-post source height increased from $\sim 3\ r_{g}$ to $\sim25\ r_{g}$ with the luminosity.

This study builds on the timing and spectral analysis of reverberating AGN reported by \cite{2022MNRAS.514.5403H}, HYC22 hereafter, using a variety of similar spectral flux levels as found in the spectra of each source. In addition we explore all suitable individual \emph{XMM-Newton} observations of 1H 0707-495 and IRAS 13224-3809 between 2000 and 2016. Our aim is to develop the model initially created by CY17 to explore these time lag signatures in an extended corona scenario. These well studied AGN are selected due to their abundance of long observations, however it should be noted that these are extreme narrow line Seyfert 1 galaxies.

\section{Observations and data reduction}
The data for all observations outlined in Table~\ref{sample_list} were downloaded from the \textit{XMM-Newton} archive and processed using standard methods. The time lag estimates calculated between the soft ($0.3 - 0.8$~keV) and hard ($1 - 4$~keV) energy bands reported by HYC22 have been used throughout this study. 

\begin{table}
\caption[The AGN sample]{\textit{XMM-Newton} observations used in this sample. The information Includes the name of the source, the Observation ID, year, exposure time and effective exposure after cleaning. The group column refers to the low, medium and high spectral flux groups. The low counts (lc) and high counts (hc) refer to those to observations containing $<5 \text{ cts s}^{-1}~\text{and} >\text{5\,cts\,s}^{-1}$ respectively.}
\begin{tabular}{ccccc}
\hline
\vspace{0.2cm}
Source & Obs ID & Year & Exp [Eff] (ks) & Group\\
\hline
{1H0707-495} & 0110890201 & 2000 & 46[41] &  Med (lc)\\
& 0148010301 & 2002 & 80[76] &  Hi (hc) \\
& 0506200201 & 2007 & 41[38] & Lo (lc)\\
& 0506200301 & & 41[39] &  Med (hc)\\ 
& 0506200401 & & 43[41] & Hi (hc)\\ 
& 0506200501 & & 47[41] &  Hi (hc)\\ 
& 0511580101 & 2008 & 124[111]  & Hi (hc) \\ 
& 0511580201 & & 124[93]  & Hi (hc)\\ 
& 0511580301 & & 123[84]  & Hi (hc)\\
& 0511580401 & & 122[81] & Hi (hc)\\
& 0653510301 & 2010 & 117[112] & Hi (hc)\\
& 0653510401 &  &128[118]  & Hi (hc)\\
& 0653510501 &  &128[93] & Hi (hc)\\
& 0653510601 &  &129[105] & Hi (hc)\\
& 0554710801 & 2011 & 98[86] & Lo (lc)\\
\hline
{IRAS13224-3809} & 0110890101 & 2002 & 64[61] &  Med (lc)\\
& 0673580101 & 2011 & 133[49] & Med (lc)\\
& 0673580201 & & 132[99] & Med (hc)\\
& 0673580301 & & 129[82] & Lo (lc)\\
& 0673580401 & & 135[113] & Med (hc)\\
& 0780560101 & 2016 & 141[141] & Med (hc)\\
& 0780561301 & & 141[127] & Med (hc)\\
& 0780561401 & & 141[126] & Med (hc)\\
& 0780561501 & & 141[126] & Med (hc)\\
& 0780561601 & & 141[137] & Med (hc)\\
& 0780561701 & & 141[123] & Med (hc)\\
& 0792180101 & & 141[123] & Med (hc)\\
& 0792180201 & & 141[129] & Med (hc)\\
& 0792180301 & & 141[129] & Lo (lc)\\
& 0792180401 & & 141[120] & Hi (hc)\\
& 0792180501 & & 138[122] & Med (lc)\\
& 0792180601 & & 136[122]& Hi (hc)\\
\hline
\end{tabular}
\label{sample_list}
\end{table}

We initially assume 1H 0707-495 to have a black hole mass $\log M = 6.31 M_{\odot}$  \citep{2003MNRAS.343..164B} and redshift $z = 0.0411$ \citep{1999ApJS..125..317L}. The luminosity distance from the NED database is 187\,Mpc.  This AGN has been well documented to be dominated by relativistically blurred reflection at either a low or a moderate inclination angle \citep[see e.g.][]{2009Natur.459..540F,2010MNRAS.401.2419Z,2012MNRAS.422.1914D,2013MNRAS.428.2795K,2018MNRAS.480.2650C}. We employ the inclination angle $i = 53^\circ$ as derived from the emissivity profile by \cite{Wilkins2011}. For IRAS 13224-3809 we assume a black hole mass of $\log M = 6.30 M_{\odot}$  \citep{10.1093/mnras/sty2527, alston_dynamic_2020, 2020MNRAS.498.3184C}. We also adopt appropriate values for redshift $z = 0.0406$ and inclination $i = 64^\circ$ as reported by \cite{2013MNRAS.429.2917F}. The luminosity distance from the NED database is 310 Mpc.

\section{THE EXTENDED CORONA MODEL (ECM)}

\begin{figure}
    \centerline{
        \includegraphics*[width=0.4\textwidth]{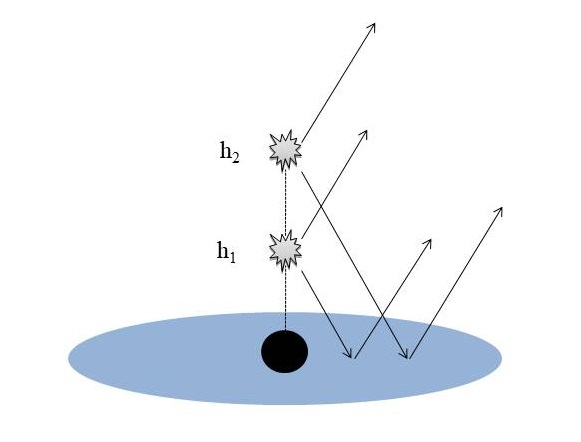}
    }
    \caption{A simple sketch of the extended corona scenario showing the two X-ray sources located on the rotation axis of an accreting black hole. The arrows show the trajectories of the continuum and accretion disc reflected photons.}
    \label{2-blobs_geometry}
\end{figure}

The extended corona model (ECM) assumes a standard geometrically thin, optically thick accretion disc \citep{1973A&A....24..337S} which extends from the innermost stable circular orbit (ISCO), or the radius of marginal stability $r_{\text{ms}}$, to 400~$r_g$ around a central black hole. Note that accreting supermassive black holes were mostly found to be rapidly spinning \citep{2021ARA&A..59..117R}, therefore to limit the number of free parameters and to avoid the model degeneracy, we fix the black hole spin to be $a=0.998$. The accretion disc is illuminated by two compact X-ray sources located on the symmetry axis at heights $h_1$ and $h_2$ which are the lower and upper source heights respectively whose amplitudes as a function of time are $x_1(t)$ and $x_2(t)$, respectively, as in CY17. We maintain the two-source to investigate the \textit{extent} of the corona and a basic sketch of this scenario is shown in Figure \ref{2-blobs_geometry}. Physically, the lower X-ray source may represent the base of a jet-like structure or the lower region of a compact corona \citep{Wilkins2015} and the upper source represents the farthest region from where a flare or response from the disc is detected hence it is interpreted as the upper extreme of the corona. A further plausible explanation of this region could be due to a periodic vertical collimation of the corona as a jet launching event subsides \citep{Wilkins2015a}. These explanations also suggest that photon emission is beamed vertically away from the disc.

The photon trajectories were traced along Kerr geodesics as described by \cite{1972ApJ...178..347B} and essentially outlined by CY17 by first considering the flares of the two X-ray sources as two separate delta functions and tracing the photons between the two sources, the disc and the observer along Kerr geodesics. The full relativistic effects outlined by \cite{1975ApJ...202..788C} are included. The X-ray reprocessing is modelled using \texttt{REFLIONX} \citep{1991MNRAS.249..352G,1999MNRAS.306..461R,2005MNRAS.358..211R}. In CY17, many of the free parameters were reduced by making model assumptions based on the physical environment of PG1244+026, for example, the inclination angle, photon index and ionisation were fixed for simplicity at the values suggested in previous literature \citep[e.g.][]{2014MNRAS.439L..26K}. Here, the inclination angle is fixed at $i = 53^\circ$ for 1H 0707-495 \citep{Wilkins2011} and $i = 64^\circ$ for IRAS 13224-3809 \citep{2013MNRAS.429.2917F}, as described in Section 2, but both photon index and ionisation are allowed to be free. Note that a high inclination of $i > 60^\circ$ for IRAS 13224-3809 was also supported by the broadband spectral fitting \citep{Jiang2018} and simultaneous lag-frequency spectral fitting \citep{alston_dynamic_2020}. Fixing the inclination can help to avoid degeneracies in the model. Then, the model contains an ionisation parameter
\begin{equation}
    \xi(r,\phi) = 4\pi F_t(r,\phi)/n(r)
\end{equation}
where $F_t(r,\phi)$ is the total flux due to both X-ray sources per unit area of the disc at $(r,\phi)$, and $n(r)$ is the disc density, $n(r) \propto r^{-p}$.

The aim is to reproduce the high frequency soft (reverberation) lags and the harder lags seen at lower frequencies which are associated with propagating fluctuations in the disc. \cite{2001MNRAS.327..799K} and \cite{2006MNRAS.367..801A} describe the low frequency hard lags by mass accretion fluctuations propagating inwards through the disc where the central region contains a source of harder X-rays. That is, the fluctuations modulate soft X-rays first resulting with the hard lags, so the soft bands are dominant first before being overcome by the hard band. Assuming these fluctuations cause the central X-ray sources to respond at different times, the extended corona scenario models this source response using a cut off power law
\begin{equation}
    \Psi_i(t) \propto t^{-q_i} \text{exp}(-t/t_\text{max})
\end{equation}
where the subscripts $i=1$ and 2 refer to the parameters of the lower and upper sources, respectively. How the function decays with respect to time is determined by $q_i$, where $t = 0$ and $t_\text{max}$ is the beginning and the end of the source response. Only the time difference between the two responses is relevant.

In addition, the model makes use of the parameter $t_\text{shift}$ to delay the response of the second source that reacts slower to primary variations of the first source. In essence, these inward propagating fluctuations can produce primary variations in the disc that will affect the flux of the lower X-ray source at time $t$ = 1 and propagate upwards to the upper X-ray source taking time $t_\text{shift}$. While the disc response usually obtained from the ray-tracing simulation is a function of energy and time, the source response is independent of energy. However, the source variability remains dependent of energy via $F_i(E) \propto E^{-\Gamma_i}$, where $\Gamma_i$ is the photon index of the X-ray continuum of the $i^{\rm th}$ source. The source variability in the energy band $E_j$ is
\begin{equation}
x_i(E_j,t)~\propto~F_i(E_j)x_0(t) \otimes \Psi_i(t),
\end{equation}
so that the lower and upper sources respond in different ways to the primary variations $x_0(t)$, due to $F_i(E_j)$ and $\Psi_i(t)$. We initially set $\xi$ for a neutral to highly ionised environment where $\log \xi = 0.0 - 3.0$ respectively and allow to vary during the fitting procedure. Other initial assumptions are made by setting the Fe abundance to solar and fixing the decay of the lower-source response $q_1 = 0.5$.

We began by creating the disc model by integrating photon paths from the observer to the disc assuming a maximum black hole spin where $a = 0.998$ with the disc extending from ISCO out to $400 r_g$. The next task was to integrate photon paths from a single X-ray source to the disc, after which the spectra was calculated using \texttt{REFLIONX} for a given source height and disc inclination already computed in the previous steps. The final step was to calculate the reverberation signatures by looking at the already computed source-to-disc and observer-to-disc ray tracing runs. The disc response function, $\psi_i(E_j,i)$, was computed and the variability of the disc reflection can be calculated via a convolution term: $x_i(E_j,t) \otimes \psi_i(E_j,i)$. 
The observed X-ray variability then can be written as the sum of the source and the disc variability. The lag-frequency spectra were calculated from the Fourier-phase differences between two energy bands, following the standard techniques \citep[e.g.][]{2014MNRAS.438.2980C,2014MNRAS.439.3931E,2015MNRAS.452..333C}.

Note that the light curve in each energy band always contains both continuum and reflection components. The contamination of the continuum flux in the reflection-dominated band, and the reflection flux in the continuum-dominated band cause dilution effects which can reduce the lag amplitude, meaning that the measured time lags are shorter than the intrinsic time lags. The dilution effects modify the shape of the lag-frequency spectrum without affecting the frequency at which the lags occur. For discussions on dilution see, e.g., \cite{Uttley2014,2014MNRAS.439L..26K,2015MNRAS.452..333C}. In order to deal with dilution effects, the variability of the reflected photons of the disc is normalized using the reflected response fraction (defined as the reflected flux/continuum flux) of all energy bands $E_j$. We include a brightness parameter $b$ (as measured in the frame of the observer) as a ratio of the brightness of the lower and upper source, $b_1$ and $b_2$ respectively by defining $b = b_2/b_1$ and fix $b_1 =1$ and allow $b$ to vary between 1--3. The continuum flux of the lower source will be less than the upper source due to its closer proximity to the black hole, therefore photons will be subject to light bending effects towards the centre \citep{2004MNRAS.349.1435M}. 

\begin{figure}% PATH: /home/steff075/phd_ecm/ecm-files/plot_tmax_script.py
    \centering
    \includegraphics*[scale=0.455]{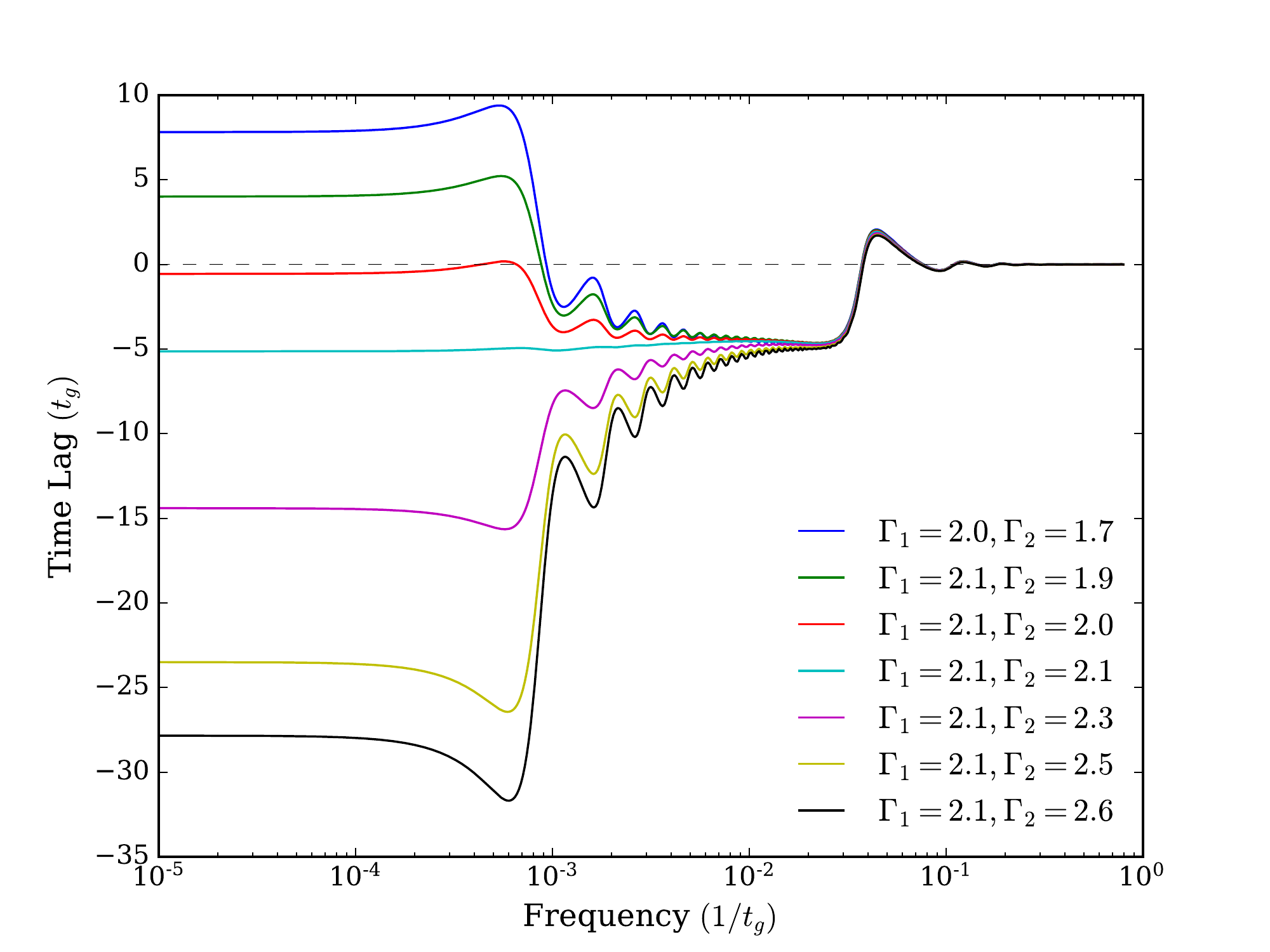} \\
    % PATH: /home/steff075/phd_ecm/ecm-files/plot_tmax_script.py
    \includegraphics*[scale=0.455]{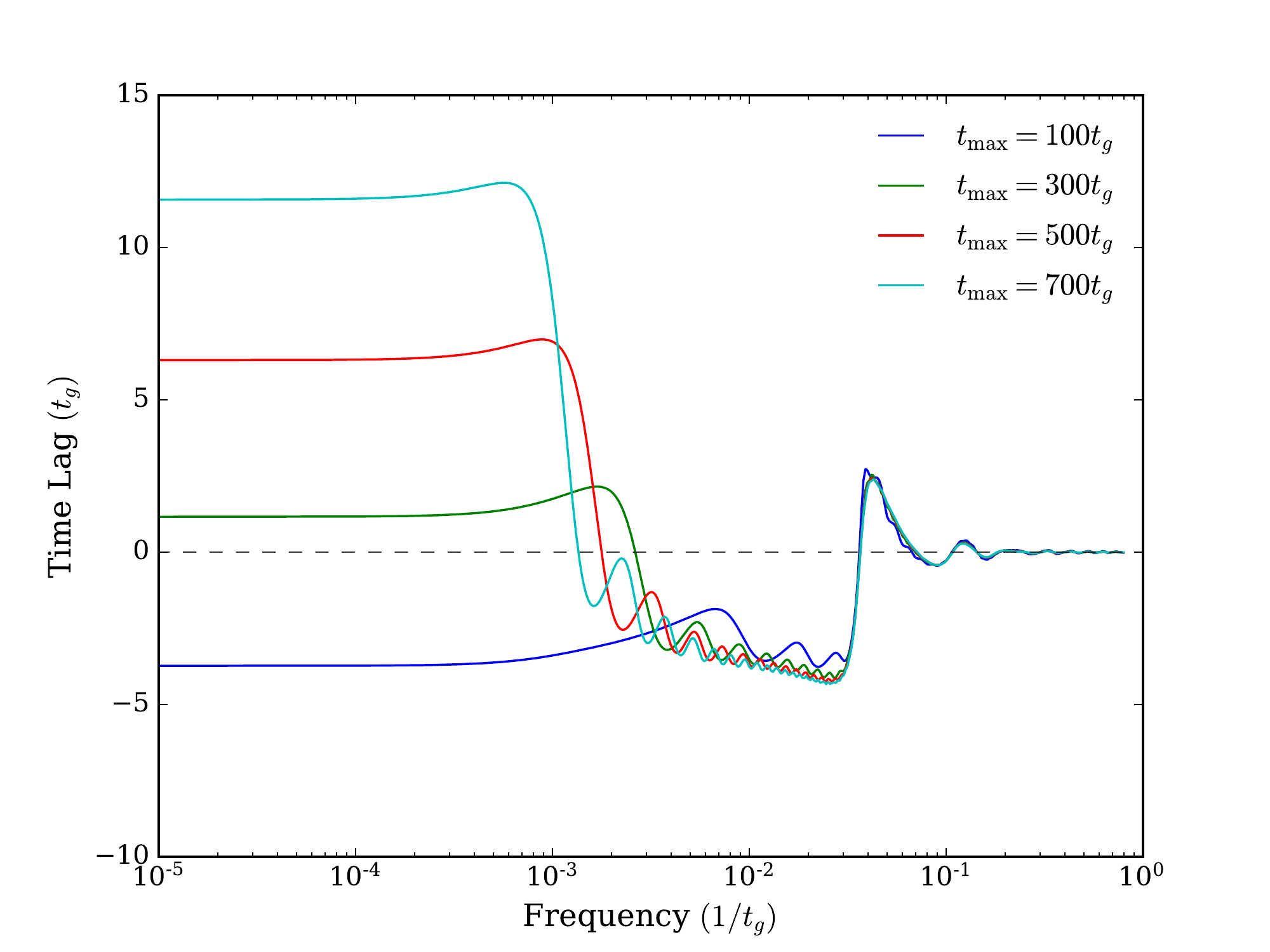}
    \caption{The frequency dependent lags varying with $\Gamma$ and $t_{\text{max}}$. Both panels were set at $i = 53^\circ$, $h_1 = 2$, $h_2 = 3$,  $Fe = 1.0$, $\log \xi = 1.0$ and $q_2 = 0.5$. The top panel shows the frequency dependent lags varying with $\Gamma_1$ and $\Gamma_2$. The positive low frequency fluctuation lags and the negative soft reverberation lags are produced when $\Gamma_1 \neq \Gamma_2$. The bottom panel shows lag behaviour with different values of $t_{\text{max}}$, where $\Gamma_1 = 2.0$ and $\Gamma_2 = 1.5$, $b = 2.0$ and $t_{\text{shift}} = 10$.}
    \label{gamma_tmax}
\end{figure}

The model is capable of reproducing the prominent time lag features seen in AGN when $\Gamma_1 \neq \Gamma_2$ as seen in the top panel of Figure ~\ref{gamma_tmax}. In addition $\Gamma_1 > \Gamma_2$ for positive low freqency lags and $\Gamma_2 > \Gamma_1$ for negaitive low frequency lags. As $t_\text{max}$ increases, so does the amplitude of the hard lags as seen in the bottom panel of Figure ~\ref{gamma_tmax} and aliasing effects (phase wrapping) are moved to lower frequencies and positive lags will switch to negative lags. The low frequency hard lags increase with energy and the softer reverberation lags dominate at higher frequencies, consistent with the `two-mechanism' features of propagation and reflection and was also consistent with the traditional spectral features that are widely explained by reflection from the inner disc.

We generated the initial model by running the code for a course range of parameter values to fit the lag-frequency spectra of the fully combined observations 1H0707-495 using an inclination angle $i = 53^\circ$. We set the density profile to constant $(p=0)$ and the response decay function $q_2 = 0.5$ and iron abundance $Z_\text{Fe} = 1.0$. The lower source height was fixed at $2\ r_g$ and upper source height was allowed to vary between $3 - 20\ r_g$ thus obtaining the \emph{extent} of the corona. While fixing the lower source at $2\ r_g$ is a pragmatic choice that simplifies the fitting and limits the number of free parameters, having one source this close to the horizon is comparable to the limits by previous spectral or timing modelling, \citep[e.g.,][]{2013MNRAS.428.2795K,2018MNRAS.480.2650C,2020MNRAS.498.3184C}. In addition, $\Gamma_1$ and $\Gamma_2$ were allowed to vary between $1.4-3.3$ in line with \texttt{REFLIONX}, the ionisation parameter specified at the ISCO could vary from $0 - 3$ for neutral and highly ionised exploration respectively. The brightness parameter was also allowed to vary from $1 - 3$. The remaining free parameters were $t_{\text{max}}$ and $t_{\text{shift}}$ and we initially set these ranging from $50 - 600\ t_g$ and $0 - 120\ t_g$, respectively. The black hole mass for each source was initially tested using sensible values from literature \citep{2003MNRAS.343..164B,alston_dynamic_2020} and allowed to vary during the fitting procedure. 

\section{Results}
\subsection{Model testing on 1H 0707-495}

\begin{figure}% TRIM={A,B,C,D} = {Left, Bottom, Right, Top}
    \centering
    % PATH /data/typhon1/reverb/1H0707-495/isis_emcee/1H0707-comb-fit-emcee.sl
    \includegraphics*[trim={3cm 0.8cm 7.1cm 4.5cm},clip,scale=0.22]{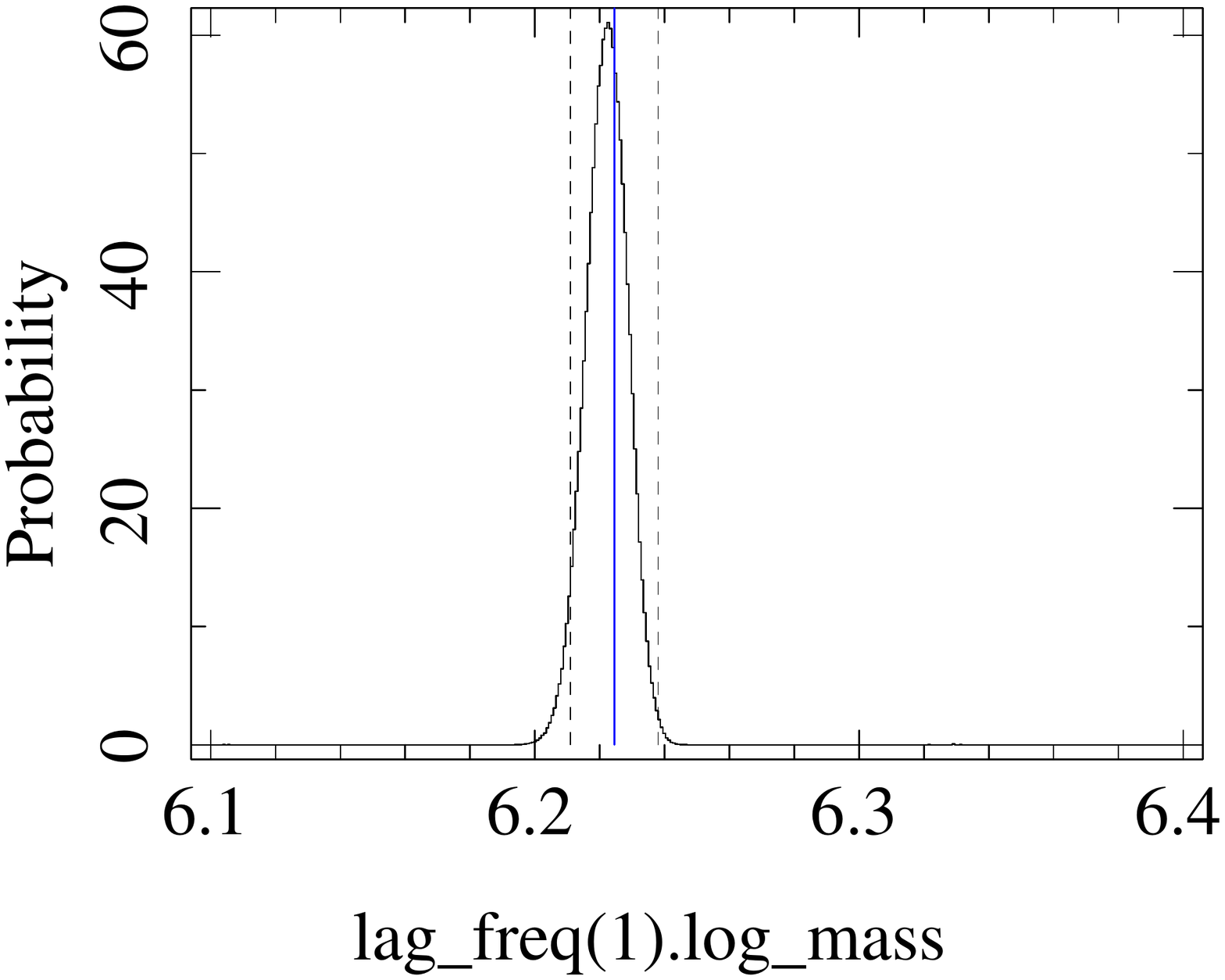}
    \put(-50,85){\small{1H 0707-495}}
    \includegraphics*[trim={2.8cm 0.8cm 7.1cm 4.5cm},clip,scale=0.22]{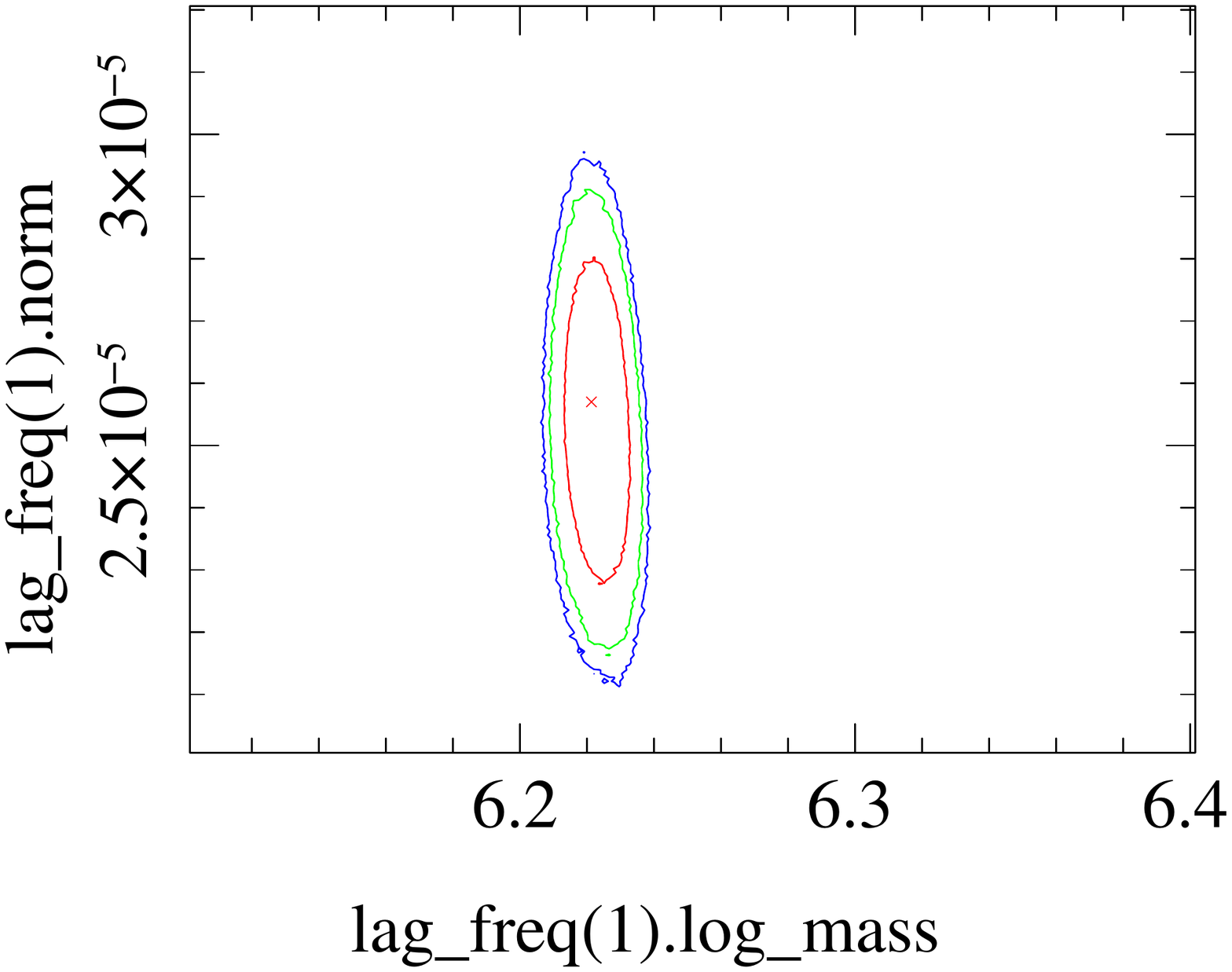}
    \put(-50,85){\small{1H 0707-495}}\\[1pt]
    % PATH /data/typhon1/reverb/IRAS13224-3809/IRAS-comb-fit-emcee.sl
    \includegraphics*[trim={3cm 0.8cm 7.1cm 3cm},clip,scale=0.22]{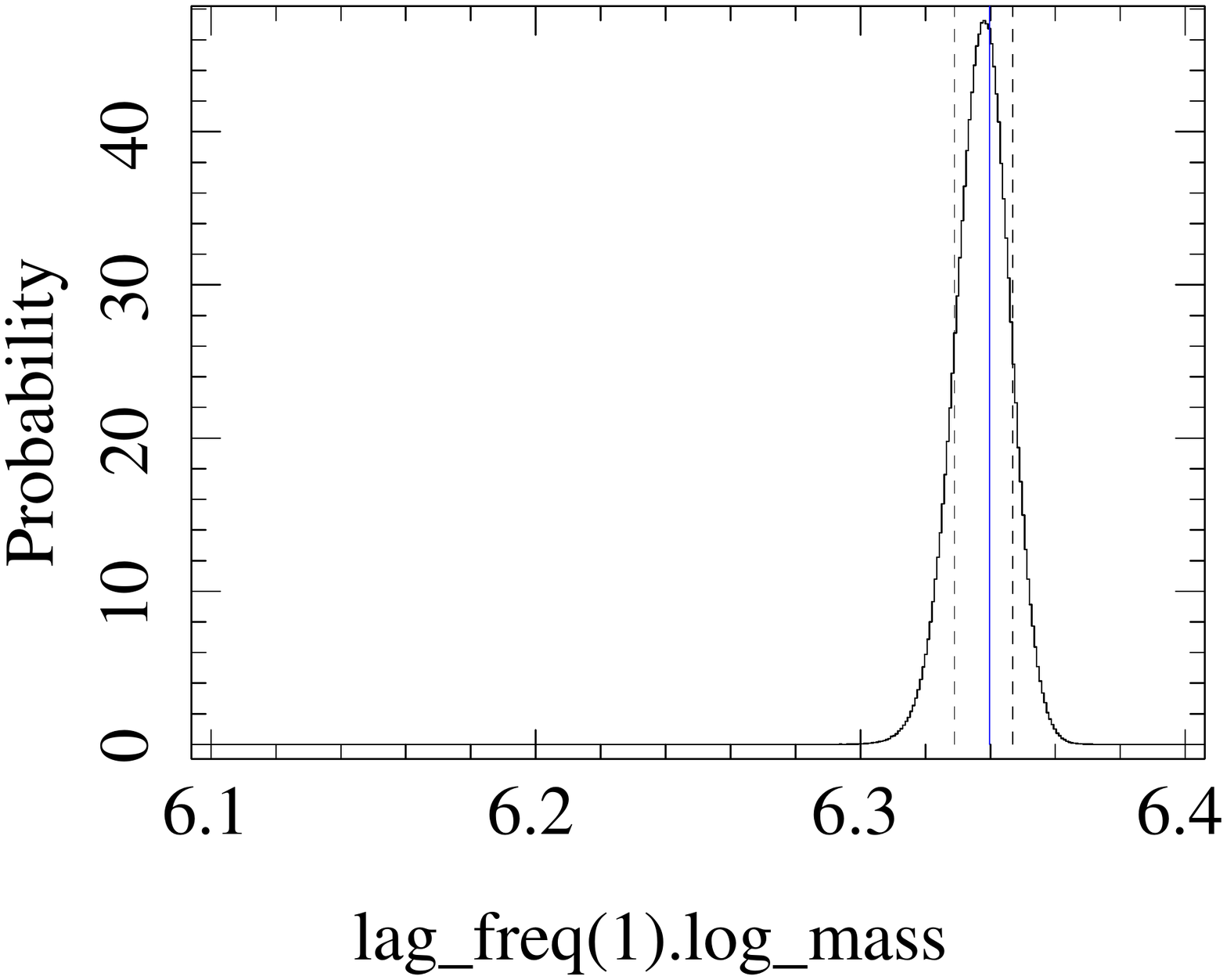}
    \put(-98,85){\small{IRAS 13224-3809}}
    \includegraphics*[trim={2.8cm 0.8cm 7.1cm 3cm},clip,scale=0.22]{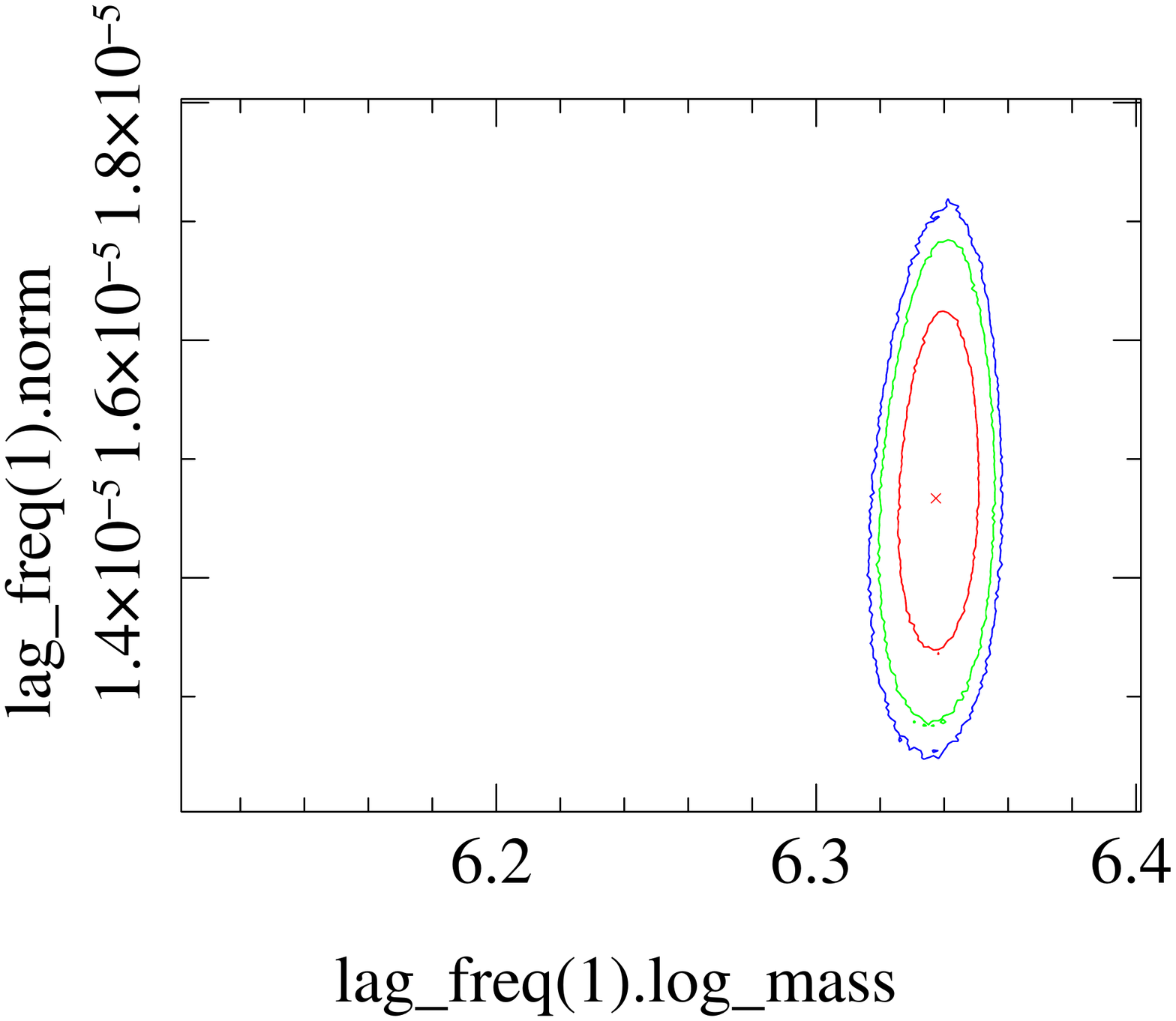}
    \put(-98,85){\small{IRAS 13224-3809}}
    \caption{Independent (emcee) tests of the posterior density for the combined observations of 1H 0707-495 \emph{(top row)} with mean mass $\mu$ = 6.22 (solid blue line) and $\sigma$ = 0.01 (black dashed lines).  IRAS 13224-3809 is also shown \emph{(bottom row)} with mean mass $\mu$ = 6.34 and $\sigma$ = 0.01. The contour plots of the mass and normalisation are shown in the right column. All $x$-axes are identical for comparison.}
    \label{fig:posterior}
\end{figure}

The ECM fitting was developed using the Interactive Spectral Interpretation System (ISIS) Version 1.6.2-27. Generally, much better fits were obtained using single observations than combining the observations into groupings. Although spectral combinations are useful for obtaining snapshots of the various epochs, the true variability is best revealed by examining each observation in turn, however this is time consuming and computationally expensive. Firstly, we independently examined the posterior distribution for the mass using combined observation best fit. For this inspection we use the \texttt{isis\_emcee} module outlined at \cite{RemeisWiki2018} by adopting the methods of the MCMC hammer \citep{2010CAMCS...5...65G,2013PASP..125..306F}. We initiate 500 walkers and run the chain with 10,000 iterations. The walkers converged tightly after 1,000 steps and settled into an acceptance rate of $\sim$0.7 after $\sim$3,000 iterations to obtain a mass of $\log(M/M_{\odot}) = 6.22 \pm 0.01$, where the errors are calculated at the 90\% confidence limit. The posterior density and contour plot is shown in the top row of Figure~\ref{fig:posterior}. This value is comparable to that reported in the literature \citep[e.g.][]{2005ApJ...618L..83Z,2010MNRAS.401.2419Z} although uncertainties in the estimations have been acknowledged. This initial model fit was reasonable where the $\chi^2$ was 2.32 and the upper source was located at $20.00^{+0.00}_{-0.11}\ r_g$. 

For this mass and geometry, the photon indices $\Gamma_1$ and $\Gamma_2$ were well constrained at $2.6^{+0.01}_{-0.07}$ and $2.1^{+0.72}_{-0.01}$ respectively, where the ionisation $\log \xi \sim 3.0$. The upper source was exactly twice as bright as the lowers source given by parameter $b \sim 2$. The source response time from disc fluctuations $t_{\text{max}} = 250.00^{+67.82}_{-0.04}\ t_g$ and the time taken for the fluctuations to propagate from the lower to upper source $t_{\text{shift}} = 20.00^{+11.05}_{-11.92}\ t_g$. Although this initial fit to the combined data is promising and can provide a good description of the data, we acknowledge the reasonably large errors returned for $t_{\text{max}}$, and the loosely constrained errors found for $t_{\text{shift}}$. In general, this is typical of the fits found for all data. 

\begin{table}
\centering
\caption[Extended Corona Model parameters]{The extended corona model parameter space for both sources. Note that $^f$ denotes fixed parameter} 
\label{ecm-initial-pars}
\begin{tabular}{ccc}
\\ \hline 
{Parameter} &  {1H 0707-495} & {IRAS 13224-3809}\\ \hline
$\log (M/M_\odot$) & $5.5-8$ & $5.5 - 8$ \\[3pt] 
$i$ ($^\circ$)  & $53^f$ & $64^f$ \\[3pt]
$a$ & $0.998^f$ & $0.998^f$\\[3pt]
$h_1$ ($r_g$) & $2^f$ & $2^f$\\[3pt]
$h_2$ ($r_g$) & $3-20$ & $3 - 20$  \\[3pt]
$\Gamma_1$ & $1.5-3.3$ & $1.5-3.3$   \\[3pt]
$\Gamma_2$ & $1.4-3.2$ & $1.4-3.2$   \\[3pt]
$Z_\text{Fe}$ (solar)  & $1^f$ & $1^f$  \\[3pt]
$\log \xi$ (ergs cm s$^{-1}$) &  $0-3$ &  $0-3$  \\[3pt]
$p$ &  $0^f$ &  $0^f$  \\[3pt]
$b$ &  $1 - 3$ &  $1 - 3$\\[3pt]
$q_2$ & $0.5^f$ & $0.5^f$ \\[3pt]
$t_\text{max}$ ($t_g$) &  $50 - 700$ &  $50 - 1500$ \\[3pt]
$t_\text{shift}$ ($t_g$) &  $0 - 120$ &  $0 - 150$\\[3pt] 
\hline
\end{tabular}
\end{table}

Further complexity was revealed when attempting to standardise the mass value for the remainder of the data to the value obtained from model fitting via the University of Bristol high performance computer \texttt{BlueCrystal}. A fixed mass of $\log(M/M_{\odot}) = 6.22$ generally returned much poorer fits and the fitting procedure remained computationally intensive.   

To develop this model further, the reverberation signatures were read into a table model that can be used to fit the data in ISIS and XSPEC. This step required the use of \texttt{astropy} and \texttt{heasp}, the latter of which is a component of the standard HEASOFT installation \citep{NASA-HEASARC}. The table model was initially generated using Python and the parameter space was given finer step sizes between parameter values to obtain a finer grid. The output model was huge (\textgreater40~GB) and the model loading time was too long for general usage, therefore the parameter space was constrained further. The final table model was 19~GB and the parameter space is outlined in Table~\ref{ecm-initial-pars}. Although this is still very cumbersome, further constraints at this stage were halted due to dynamic benefits of the model parameter space.

The first table model fit was conducted on the combined data for 1H 0707-495, achieving a very good model fit to the data where $\chi^2 = 0.50$\footnote{Note that the number of degrees of freedom depends on the binning being used, and with lighter binning we would have more degrees of freedom. We have chosen to bin the data more heavily to clearly show the time lag versus frequency with higher signal-to-noise ratio, so there are fewer data bins, which means that the number of degrees of freedom is very low. The $\chi^2$ values reported here then are close to the reduced $\chi^2$ values. This also highlights a more general problem when we apply the extended corona model while the quality of the time lags is limited. The data are however, still able to strongly constrain our models, showing which regions of parameter space provide a good description of the time lags and which can be ruled out.}. This fitting was applied to all of the combination flux groups for low, medium, high etc and the results are presented in Table~\ref{1H_TM_ECM_fits}. The combined data and model fit along with the lag-frequency spectrum file predicted by the model is presented in the top panel of Figure~\ref{fig:lag-freq-vari-M}. Of course, a full table model listing every possible integer value of parameter space would be extremely difficult to achieve given the intensive computational power required, so only the closest lag-frequency, that is the \textit{predicted reverberation signatures} are shown unbinned at full resolution by the solid grey wavy lines. 

\subsection{Model testing on IRAS 13224-3809}
\begin{figure}
\centering
% TRIM={A,B,C,D} = {Left, Bottom, Right, Top}
% eg:  [trim={0.0cm 0 1.5cm 0.8cm},clip,scale=0.6]{images/1H0707_tm_comb.pdf}
%code: /data/typhon1/reverb/table-model-ecm-fits/1H0707-495/fit-table-model-comb.sl
%code: /data/typhon1/reverb/table-model-ecm-fits/IRAS13224-3809/fit-table-model-comb.sl
\includegraphics[trim={3cm 0.8cm 7.1cm 3.8cm},clip,scale=0.4]{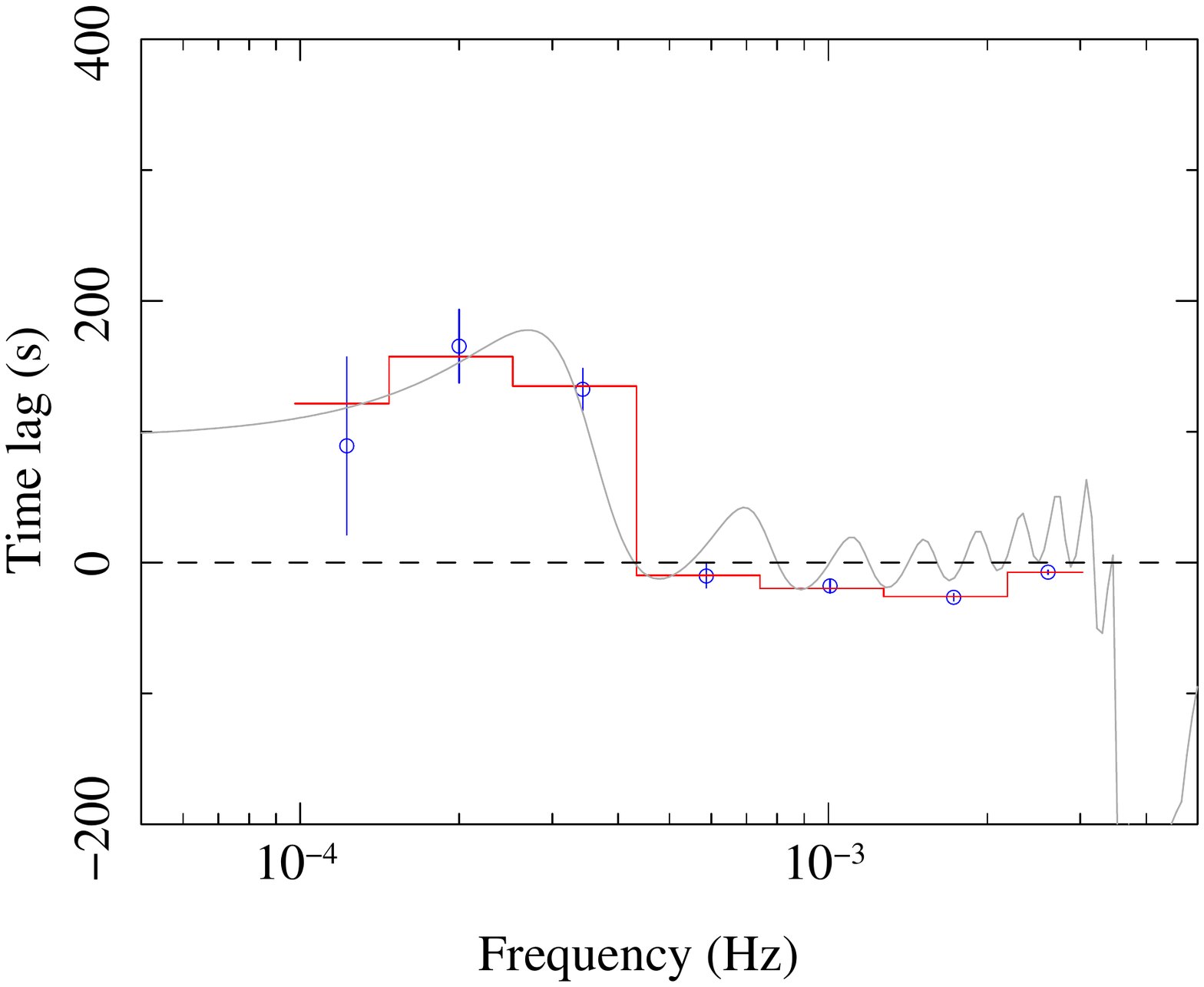} 
%\put(-150,85){1H 0707-495} 
\includegraphics[trim={3cm 0.8cm 7.1cm 3.8cm},clip,scale=0.4]{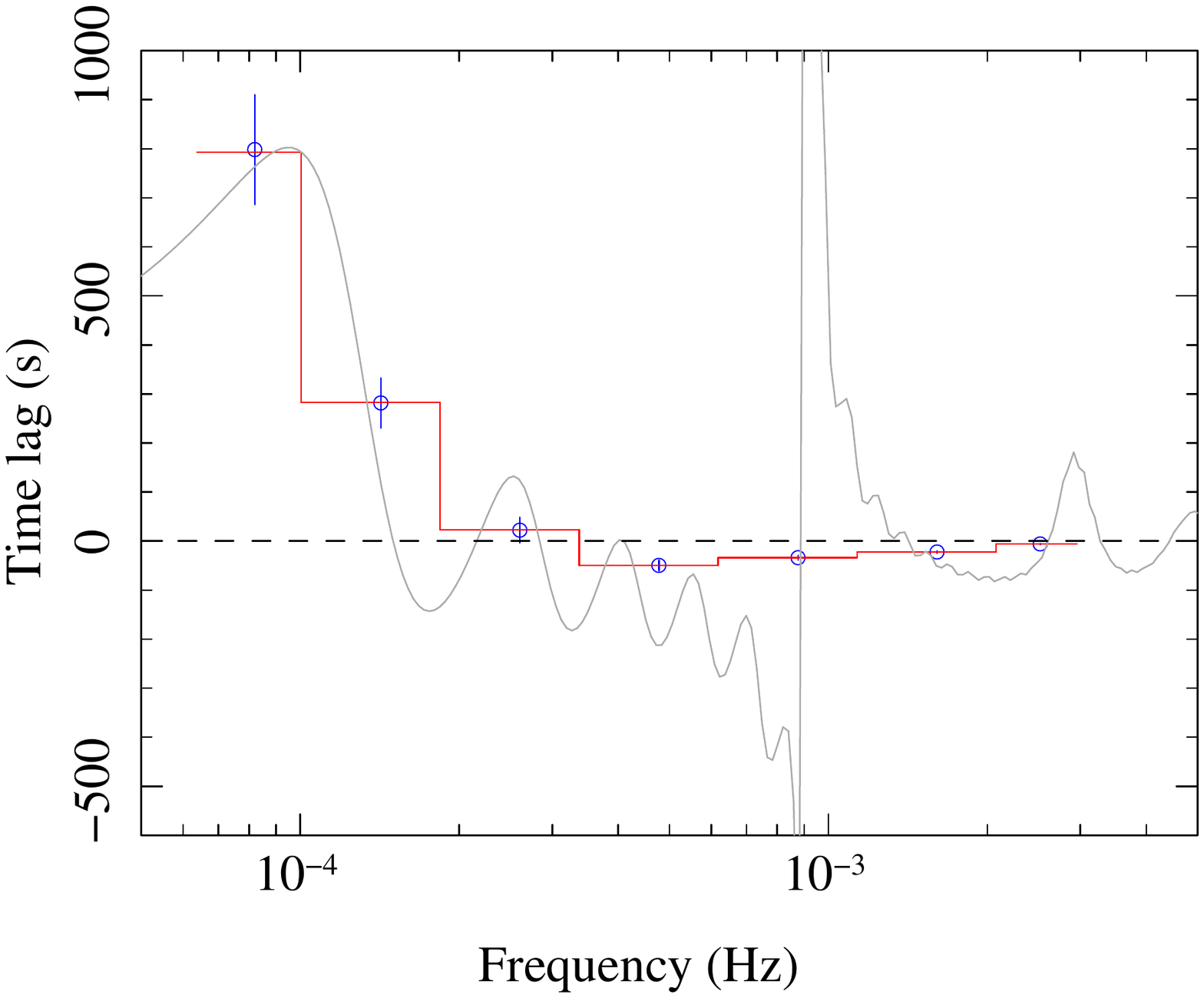}
%\put(-150,85){IRAS 13224-3809} 
\caption[ECM variable mass and TM fits] {The combined lag-frequency Table model fits for 1H 0707-495 \emph{(top panel)} and IRAS 13224-3809 \emph{(bottom panel)} showing the best model fit (red) where the mass is variable. The closest reverberation signature is shown by the grey wavy line. Note that the lags are estimated between the soft, $0.3 - 0.8$\,keV, and hard, $1 - 4$\,keV, bands. Furthermore, there is significant cancellation of positive and negative lags due to phase wrapping and binning.}
\label{fig:lag-freq-vari-M}
\end{figure}

%We assume a black hole mass of 6.8 log $M_{\odot}$ for IRAS 13224-3809 \citep{2012A&A...544A..80G} 
Once again we explore the initial model fit using \texttt{isis\_emcee} with 500 walkers and 10,000 iterations and walkers converged tightly after only $\sim200$ steps, settling into an acceptance rate of $\sim$0.7 after $\sim$300 iterations. We obtained a mass of $\log(M/M_{\odot}) = 6.34 \pm 0.01$ for IRAS 13224-3809, where the errors are calculated at the 90\% confidence limit. Again we note the model capability of returning a mass value closely comparable to $\log(M/M_{\odot}) \sim 6.3$ as reported in, e.g., \cite{alston_dynamic_2020}. The mass posterior density and contour plot, along with the model fit is shown in bottom row of Figure~\ref{fig:posterior}. The ECM fit was statistically good where $\chi^2$ = 1.47 with the upper source located at $5.00^{+5.52}_{-0.02}\ r_g$. For this mass and geometry, the photon indices $\Gamma_1$ and $\Gamma_2$ were well constrained at $2.60^{+0.02}_{-0.59}$ and $1.90\pm{0.01}$ respectively, where the ionisation $\log \xi \sim 1.0$. The upper source was about 3 times as bright as the lowers source ($b \sim 3$). The source response time from disc fluctuations $t_{\text{max}} = 1300.00^{+0.00}_{-304.36}\ t_g$. Also note that the ranges of the parameters $t_{\text{max}}$ and $t_{\text{shift}}$ required adjusting up to 1500~$t_g$ and 150 $t_g$ respectively for this source as summarised in Table \ref{ecm-initial-pars}. The model was able to provide a good statistical description of all data when fitted via the table model. The best-fit result for the combined observations is presented in the bottom panel of Figure~\ref{fig:lag-freq-vari-M}. Note that there is significant cancellation of positive and negative lags dues to phase wrapping and binning.

% Table model best fits to lag-frequency (variable M)

\subsection{Constraining the black hole mass}

\begin{figure}
\centering
% TRIM={A,B,C,D} = {Left, Bottom, Right, Top}
% eg:  [trim={0.0cm 0 1.5cm 0.8cm},clip,scale=0.6]{images/1H0707_tm_comb.pdf}
%code: /home/steff075/phd_ecm/results-spectral/plot_1H_ECM_TM_M_lag.py
%code: /home/steff075/phd_ecm/results-spectral/plot_IRAS_ECM_TM_M_lag.py
\includegraphics[trim={0.1cm 0 0 0},clip,scale=0.42]{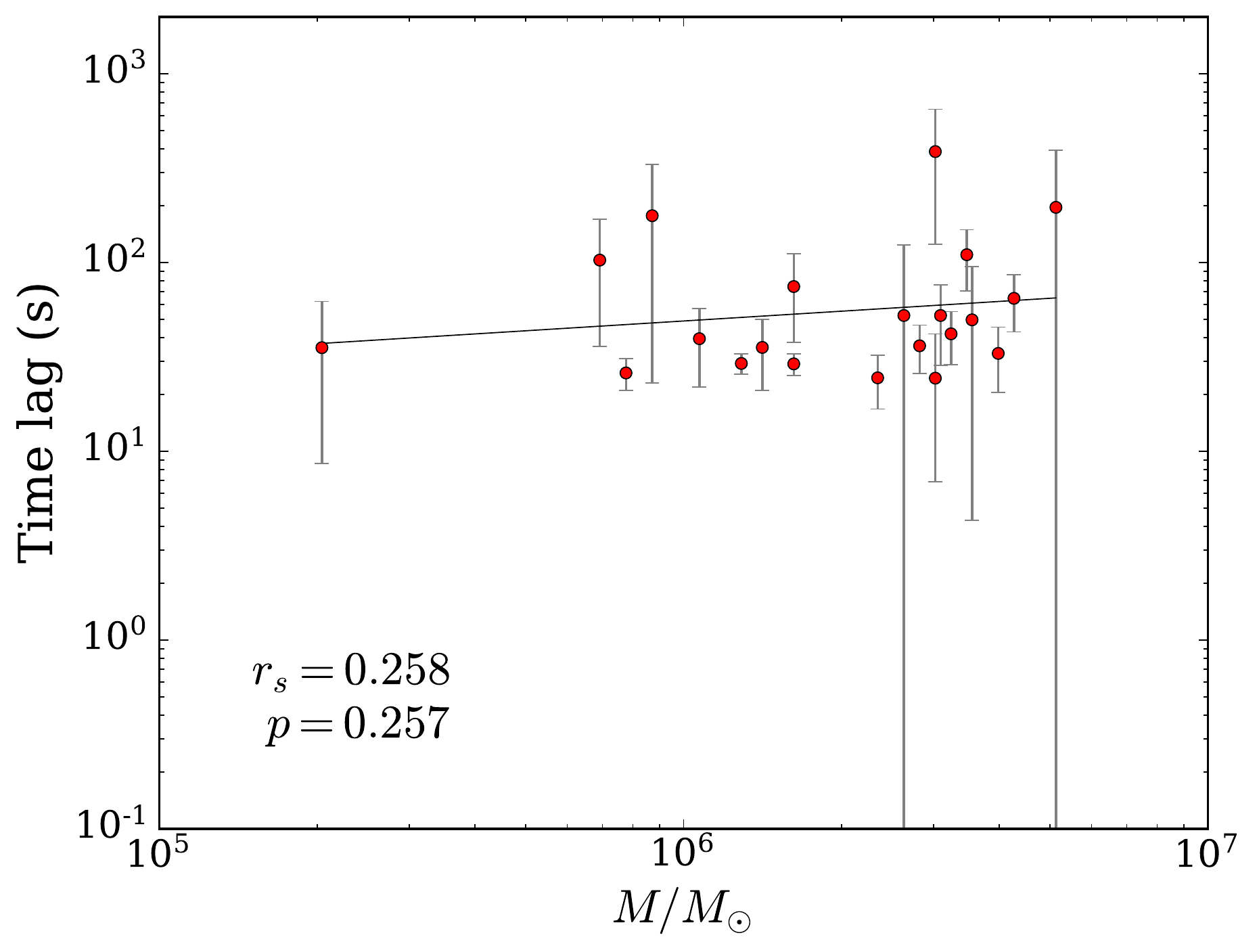} 
\put(-195,160){1H 0707-495} \\
\includegraphics[trim={0.1cm 0 0 0},clip,scale=0.42]{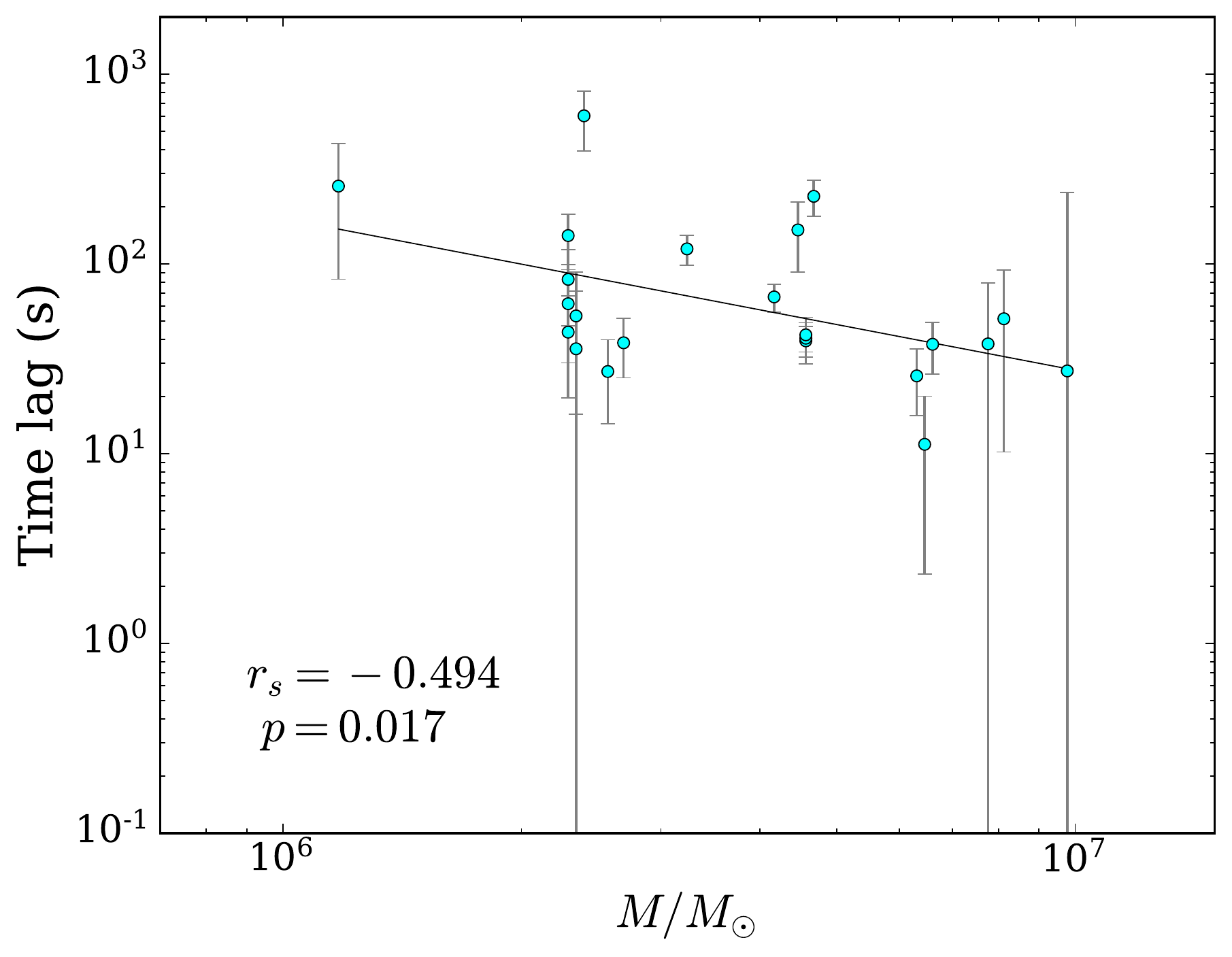}
\put(-187,160){IRAS 13224-3809} 
\caption[ECM variable mass and time lag] {The variable mass from the Table model fits as a function of the reverberation lag for 1H 0707-495 \emph{(top)} and IRAS 13224-3809 \emph{(bottom)}. Other physical processes related to the coronal properties, rather than geometry, may involve in manifesting the lag-mass scaling relation under the extended corona environment especially in IRAS 13224-3809 where the anti-correlation between the lags and the mass is seen. Of course, there is only one true value of the mass for each AGN.}
\label{fig:TM-vari-M-lag}
\end{figure}

Before exploring the simultaneously fitted ECM model, we inspected the variable mass model fits as a function of the negative (reverberation) time lags. For each source this relationship is shown in Figure~\ref{fig:TM-vari-M-lag}, suggesting that whilst the black hole mass and time lag relationship is not evident for 1H~0707-495 ($p > 0.05$), it is moderately anti-correlated in IRAS 13224-3809. Since the black hole mass for each source should be constant during these observations, the variable mass found in IRAS 13224-3809 could be induced by other geometric effects in an extended corona environment that are not related to the central mass. It could also be due to the fact that the height and the mass are degenerate \citep{2018MNRAS.480.2650C, 2020MNRAS.498.3184C}, since the mass and the source height affect the lags in a similar way. In any case, our results suggest that it might be better to fix the black hole mass when performing reverberation analysis.

\begin{figure}
\centering
% TRIM={A,B,C,D} = {Left, Bottom, Right, Top}
% eg:  [trim={0.0cm 0 1.5cm 0.8cm},clip,scale=0.6]{images/1H0707_tm_comb.pdf}
    \includegraphics[trim={3.6cm 1.8cm 7cm 4.5cm},clip,scale=0.42]{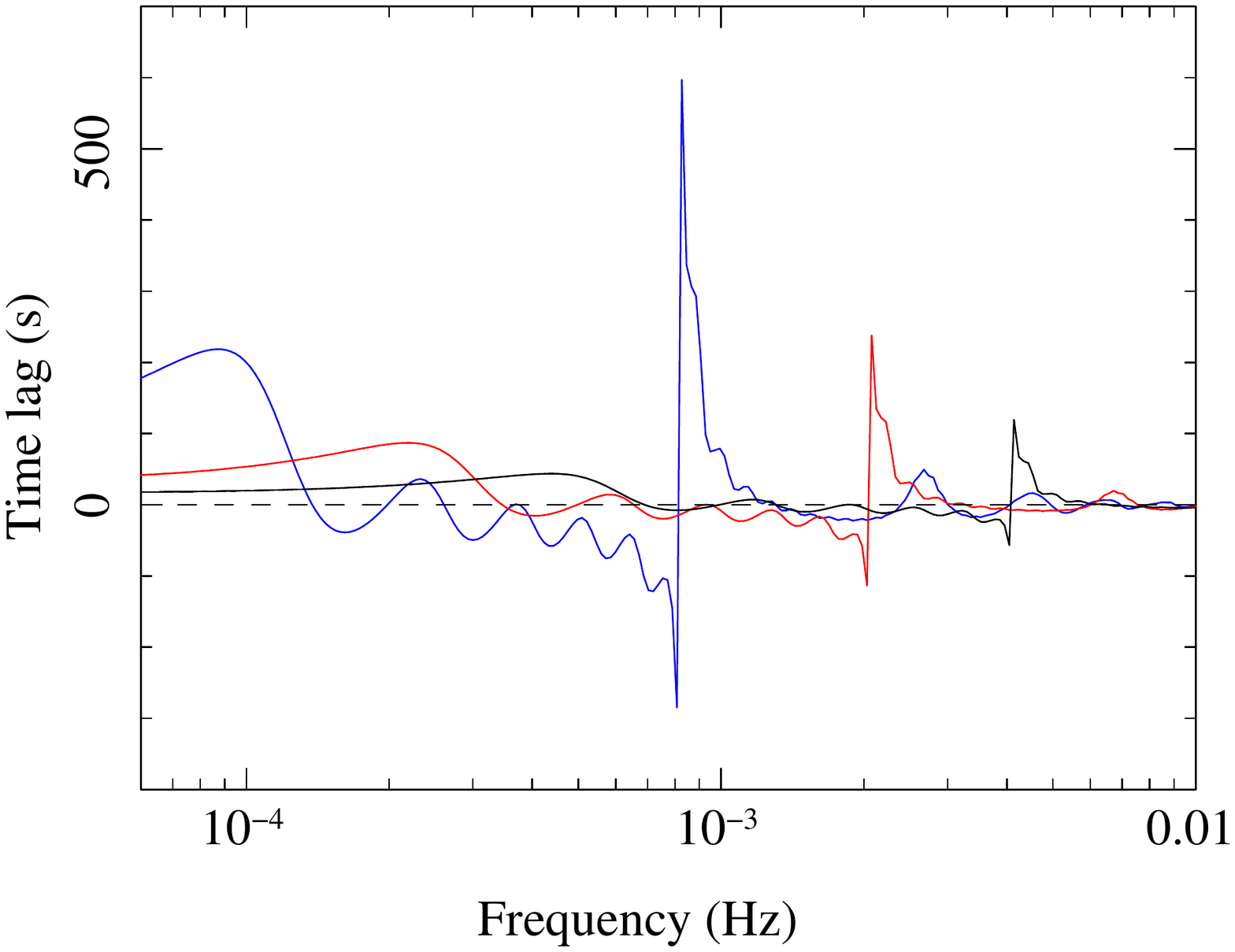}
    \put(-130,38){log$M$ = 6.7}
    \put(-100,55){log$M$ = 6.3}
    \put(-65,70){log$M$ = 6.0}\\
\includegraphics[trim={3.6cm 1.8cm 7cm 4.5cm},clip,scale=0.42]{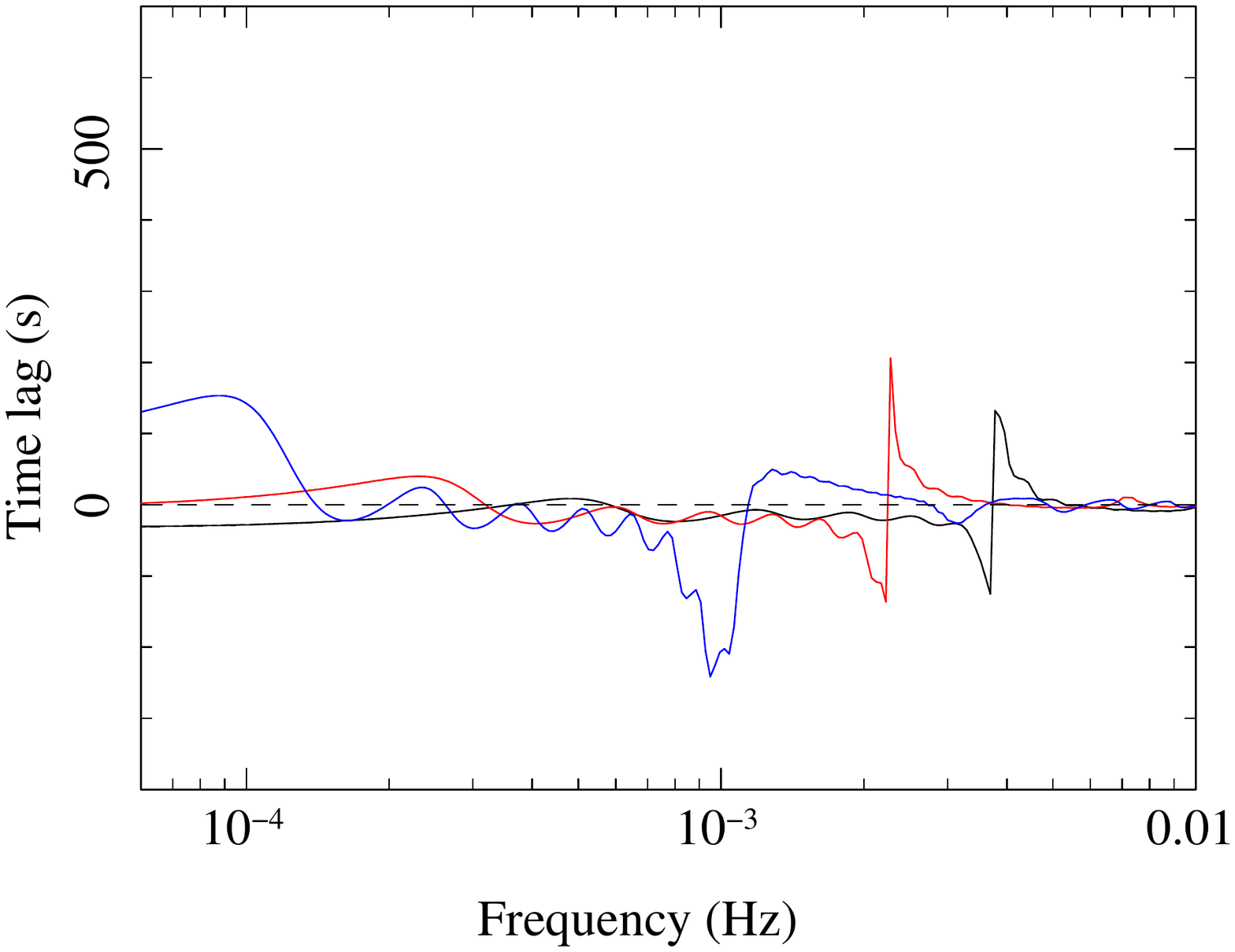}
    \put(-115,38){$h_2 = 20\ r_g$}
    \put(-80,50){$h_2 = 10\ r_g$}
    \put(-60,62){$h_2 = 5\ r_g$}    
%code: /data/typhon1/reverb/table-model-ecm-fits/IRAS13224-3809/fit-table-model-comb-MASS-CHANGE.sl
\caption[Variable mass model behaviour] {\emph{Top panel}: The variable mass behaviour for IRAS 13224-3809 combined table model where $h_2 = 13.0\ r_g$ and log$M$ = 6.0 (black), 6.3 (red) and 6.7 (blue). Specific model parameters of the reverberation signature are also provided. \emph{Bottom panel}: The variable height behaviour is shown for the same signature parameters and mass values as seen in the top panel, where $h_2 = 5\ r_g \text{(black)}, 10\ r_g (\text{red})~\text{and}~20\ r_g \text{(blue)}$. The other model parameters were  $\Gamma_1 = 2.6, \Gamma_2 = 1.7,\log \xi = 3.0, b = 2.0$, $t_{\text{max}} = 300, t_{\text{shift}} = 20$ (for both panels). Note that in the ECM model, the mass scales not only the negative reverberation lags but also the positive lags.}
\label{fig:IRAS_comb_vari_Mass}
\end{figure}

To obtain the best mass value for all data, we use similar procedures to \cite{alston_dynamic_2020} by simultaneously fitting the data. We allowed the mass to vary between values of $\log(M/M_{\odot}) = 5.5 - 8.0$ and loaded the first data set and ran the table model fit, then added the second data set where the identical parameters were tied for each set. Essentially we leave the first mass as a free parameter and subsequently tie the remaining data masses to the first data set. This method of building one large parameter file was computationally intensive. The results confirmed our expectations that not all data sets would obtain individually excellent fits and, for 1H 0707-495, the $\chi^2$ statistics ranged from $0.19 - 2.98$. This method, however, helped to reduced the large errors for $t_{\text{max}}$ and $t_{\text{shift}}$ for the majority of the data. The single mass value obtained for 1H 0707-495 was $\log(M/M_{\odot}) = 6.04\pm0.01$. The 1H 0707-495 mass constrained here is slightly smaller than the mass of $\log(M/M_{\odot}) = 6.37$ as reported in \cite{2005ApJ...618L..83Z}. In fact, \cite{2018MNRAS.480.2650C} also adopted the mass from \cite{2005ApJ...618L..83Z} and showed that if the mass is larger than this, the lamp-post model will not be able to fit the reverberation time-lag data. Our obtained mass for 1H 0707-495 is then still in the acceptable regime found by \cite{2018MNRAS.480.2650C}. We repeated these methods to obtain the single mass value for IRAS 13224-3809. All 17 data sets were reasonably well constrained statistically where $\chi^2$ ranged 0.17 -- 2.75, achieving a single mass value of log$(M/M_{\odot}) = 6.30\pm0.01$, which is closely consistent with what obtained by \cite{alston_dynamic_2020}. Figure~\ref{fig:IRAS_comb_vari_Mass} shows an overview of the behaviour of the time lags as a function of frequency for different black hole mass values and upper X-ray source locations. While the positive lags in previous literature were usually modelled independently using a power-law function, the ECM model can simultaneously produce both negative and positive lags.

\begin{comment}
The discrepancy between the mass measurements of the combined data sets, the individual data sets, and the simultaneous fitting of the individual data sets demonstrates that the data do have the statistical power to measure the black hole mass albeit subject, at present, to systematic modelling uncertainties.
\end{comment}

\subsection{ECM correlations}

\begin{figure}
\centering
\includegraphics[trim={0.1cm 0.2cm 1.5cm 0.2cm},clip,scale=0.30]{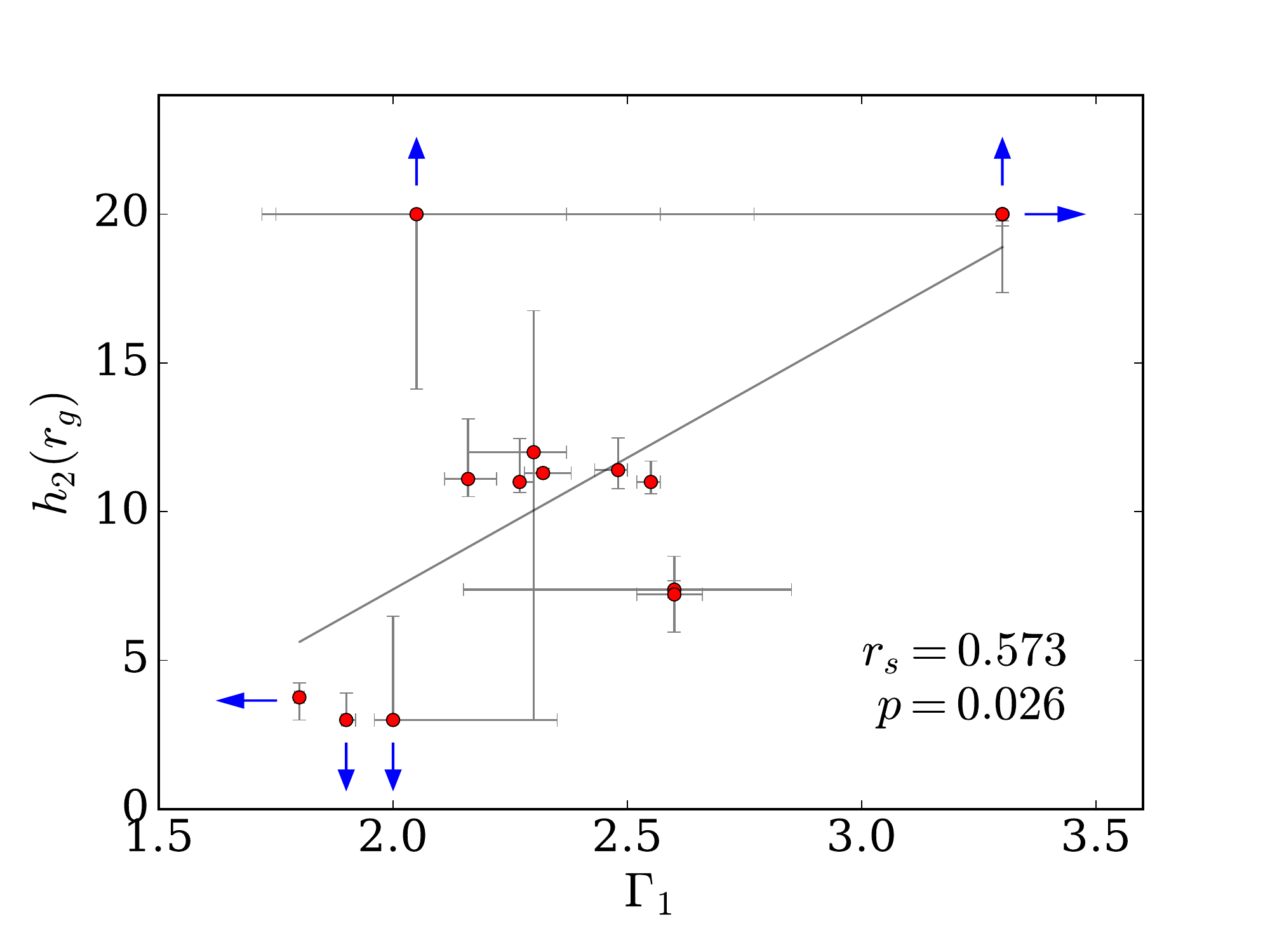}
% /home/steff075/phd_ecm/results-spectral/plot_1H_ECM_g1.py
\includegraphics[trim={0.1cm 0.2cm 1.5cm 0.2cm},clip,scale=0.30]{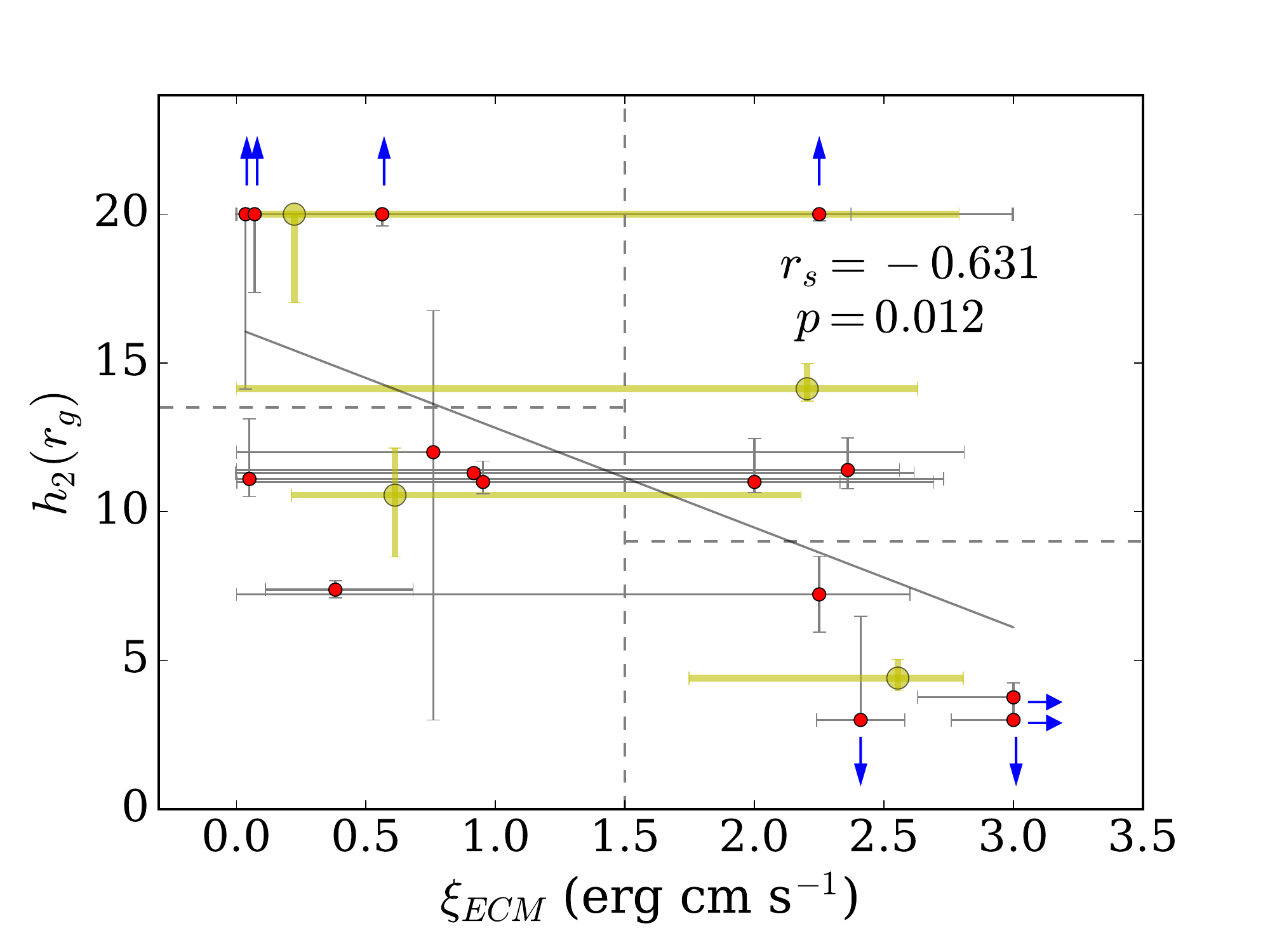} %plot_1H_ECM_h2_xi_VIS.py
\includegraphics[trim={0.1cm 0.2cm 1.5cm 0.2cm},clip,scale=0.30]{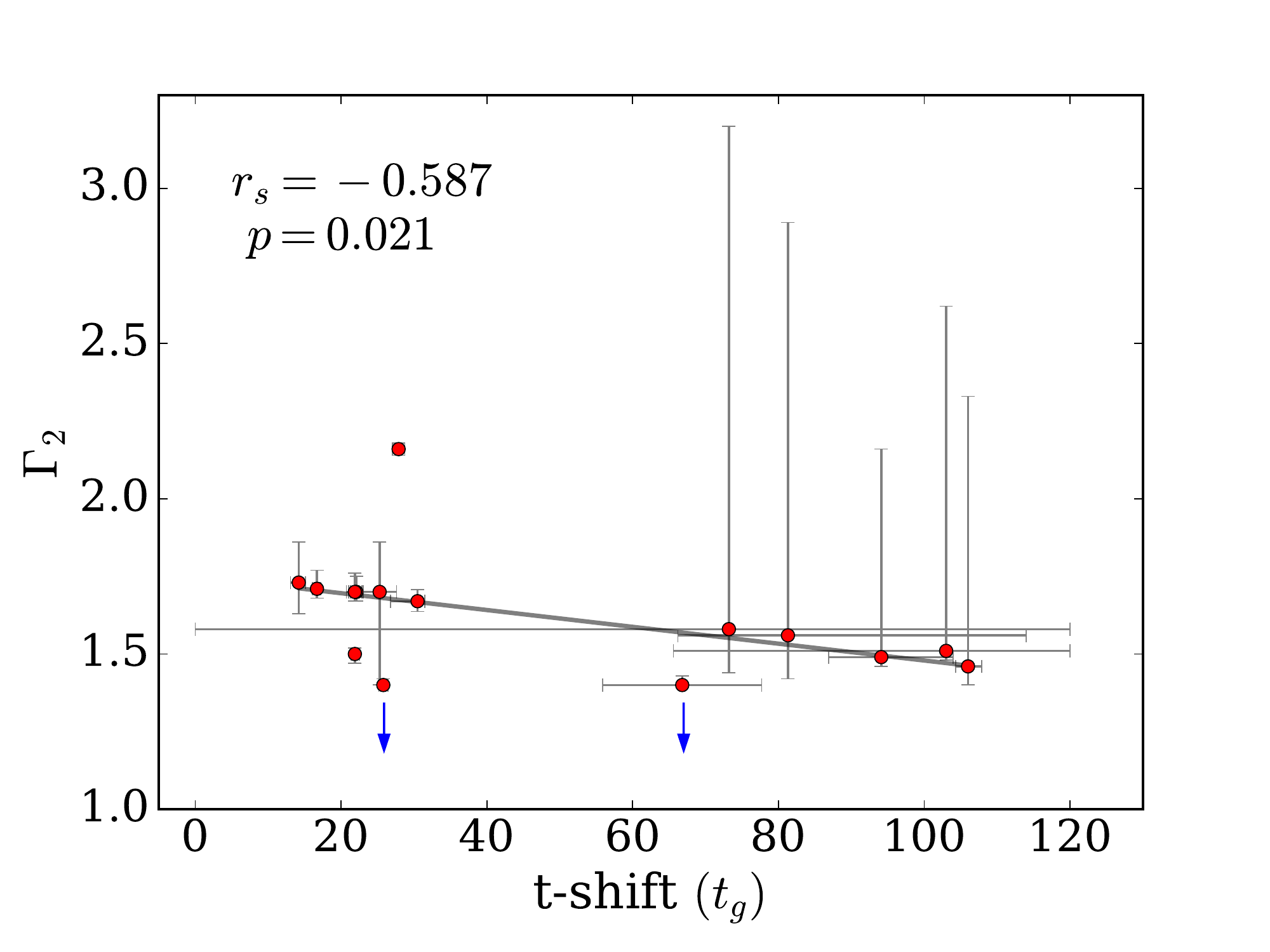}\\ %plot_1H_ECM_g2_tshift.py
\includegraphics[trim={0.1cm 0.2cm 1.5cm 0.2cm},clip,scale=0.30]{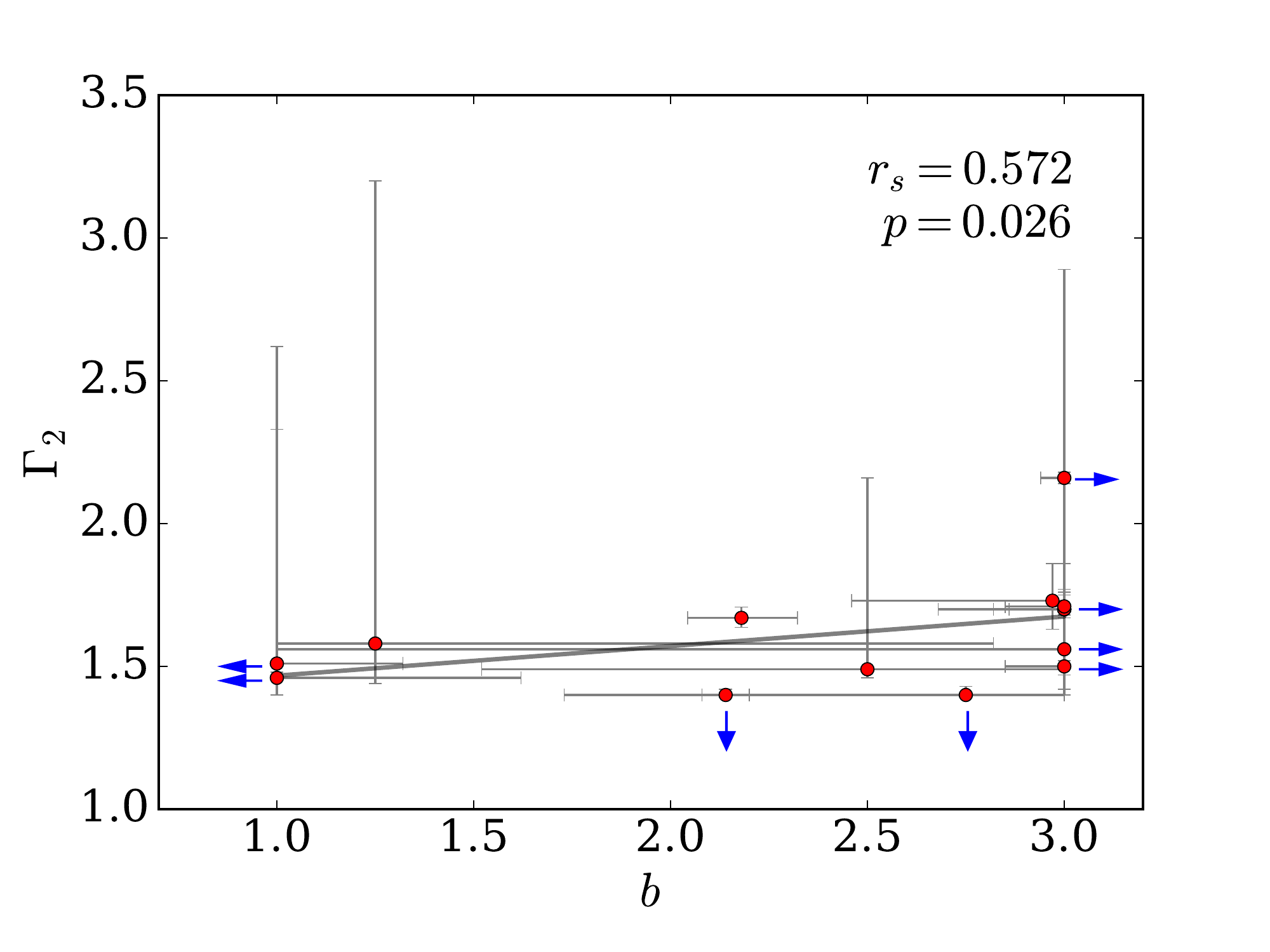} %plot_1H_ECM_g2_b.py
\includegraphics[trim={0.1cm 0.2cm 1.5cm 0.2cm},clip,scale=0.30]{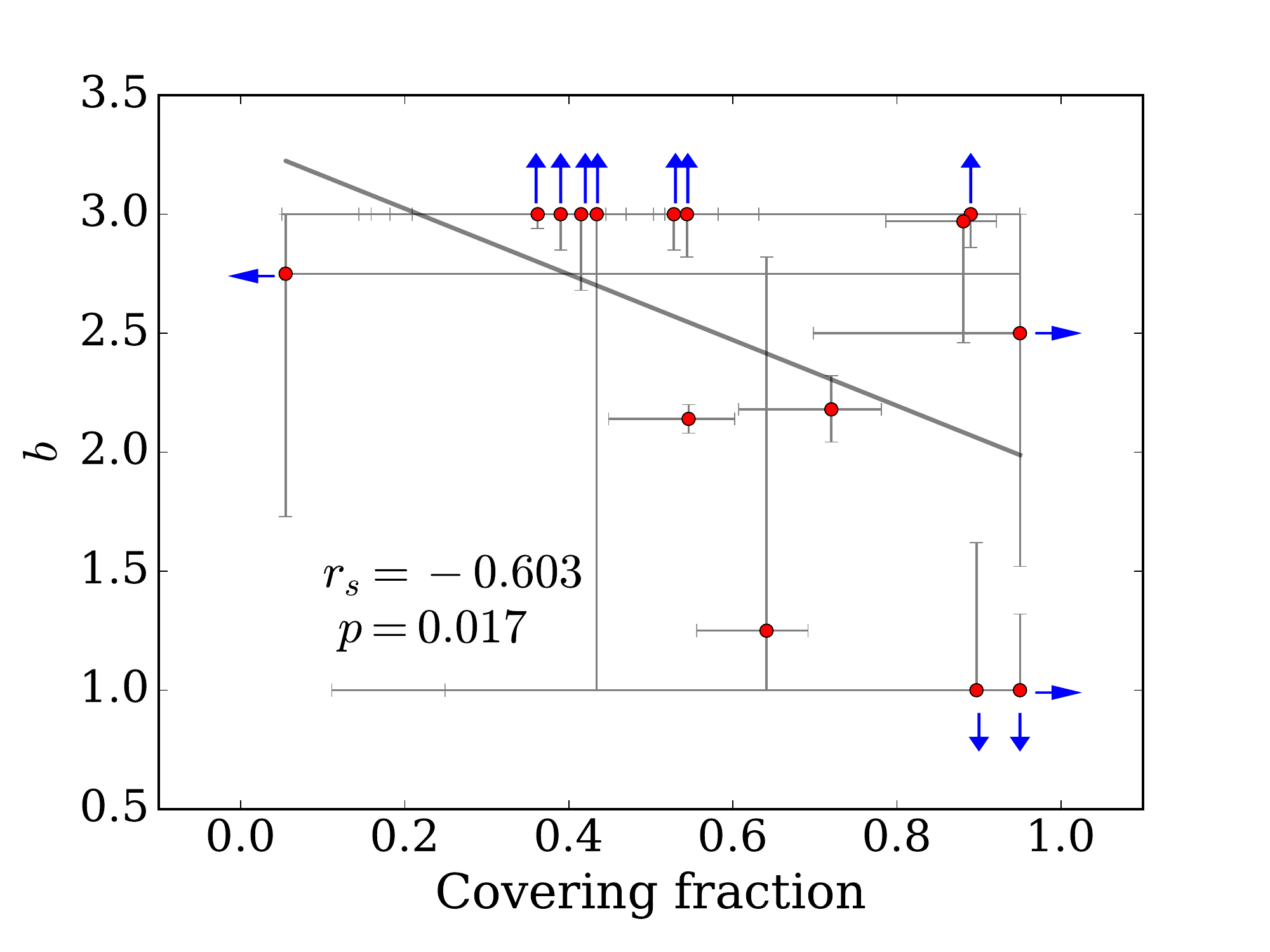} %plot_1H_ECM_b_Cvr.py
\includegraphics[trim={0.1cm 0.2cm 1.5cm 0.2cm},clip,scale=0.30]{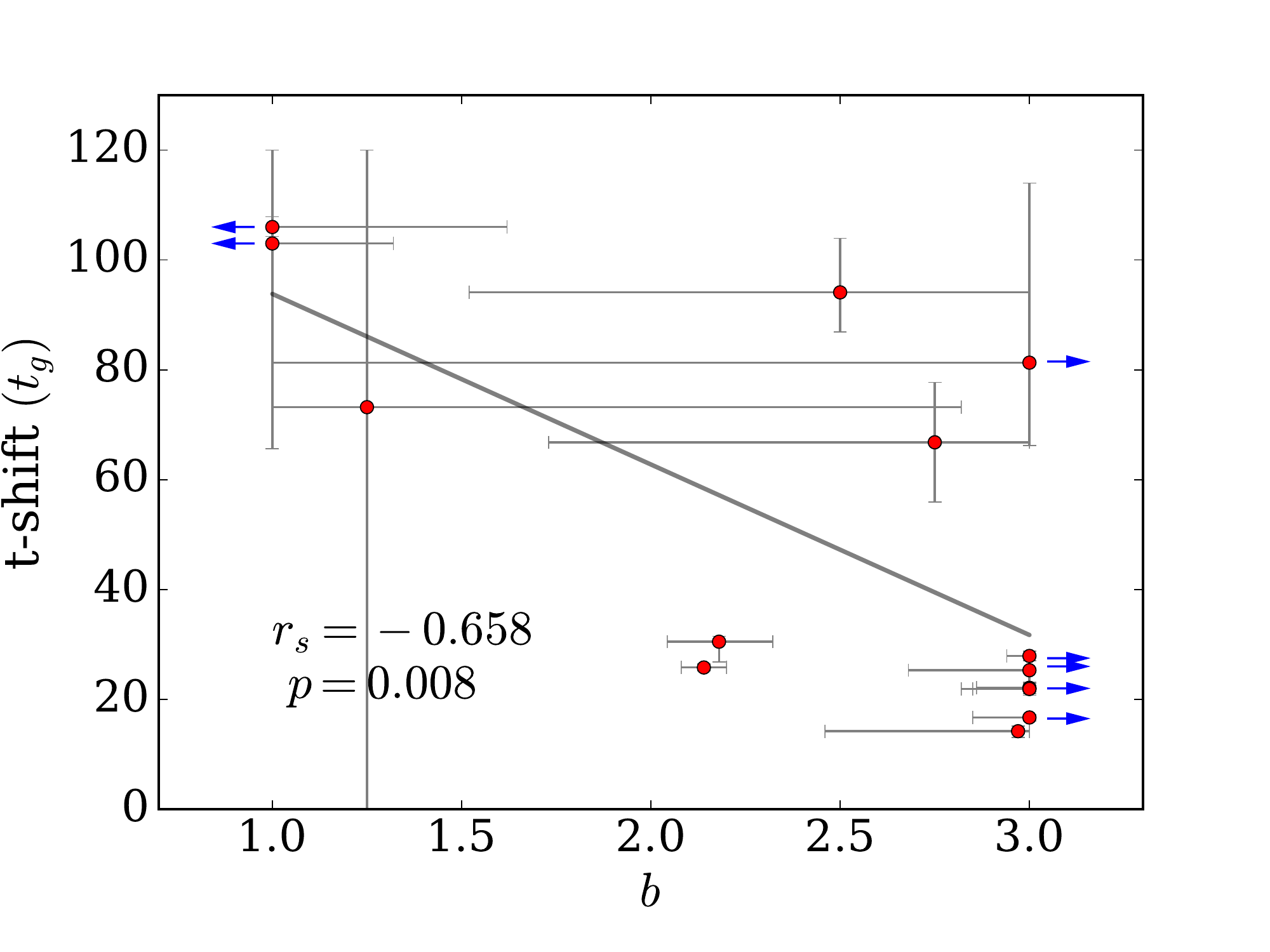} %plot_1H_ECM_tshift_b.py
\caption[1H 0707-495 correlations]{The 1H 0707-495 ECM results showing moderate to strong correlations where the $p$-value \textless{0.05}. For clarity of the $h_2 - \xi_{\text{ECM}}$ correlations \emph{(upper middle panel)} have been heavily averaged into 4 groups and shown by the yellow points for data points located in quadrants identified by the grey dashed lines.}
\label{fig:1H0707-SR-results}
\end{figure}

\begin{figure}
\centering
\includegraphics[trim={0.2cm 0.2cm 1.5cm 0.2cm},clip,scale=0.30]{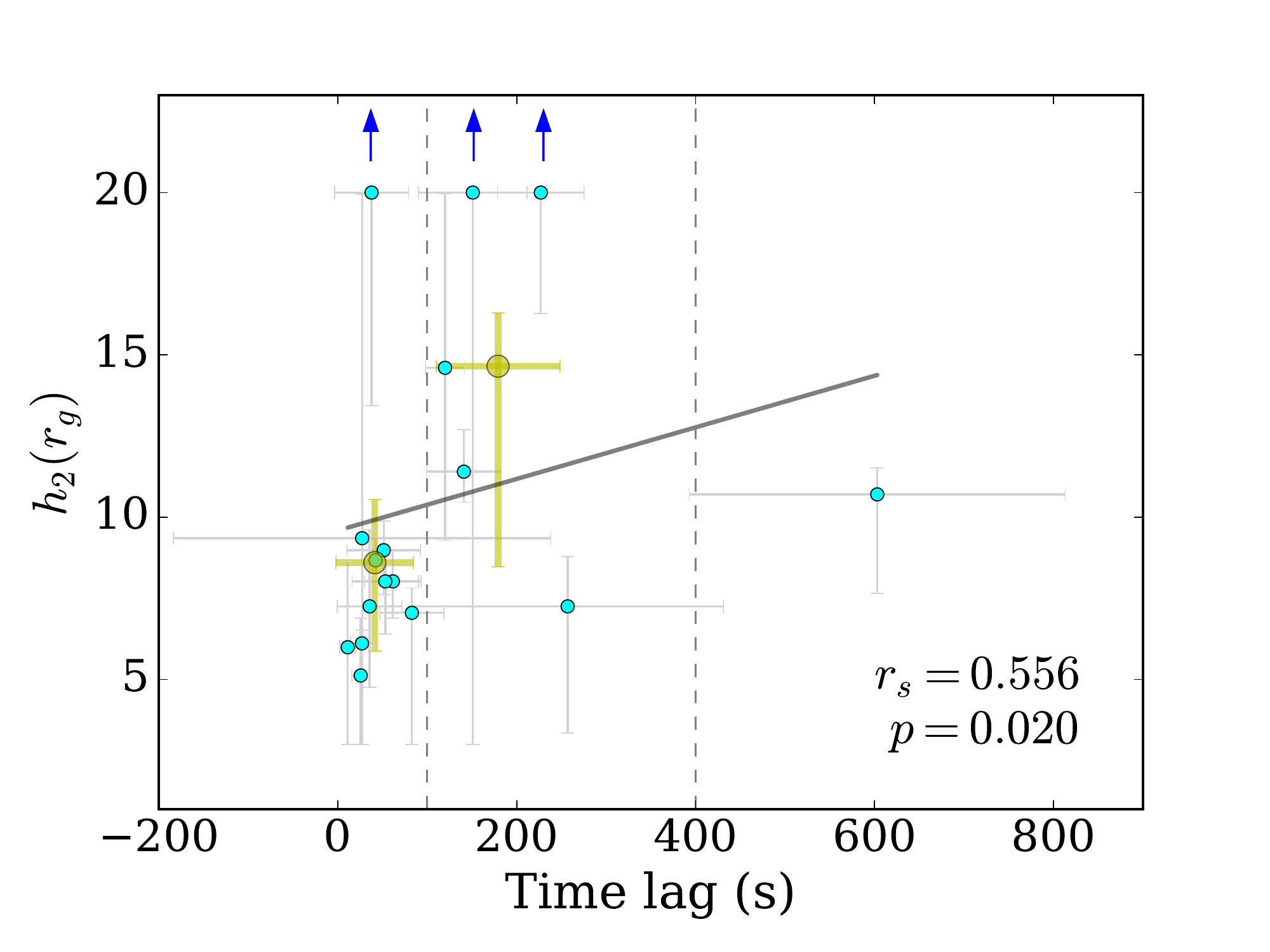}
%plot_IRAS_ECM_h2_lag_VIS.py
\includegraphics[trim={0.2cm 0.2cm 1.5cm 0.2cm},clip,scale=0.30]{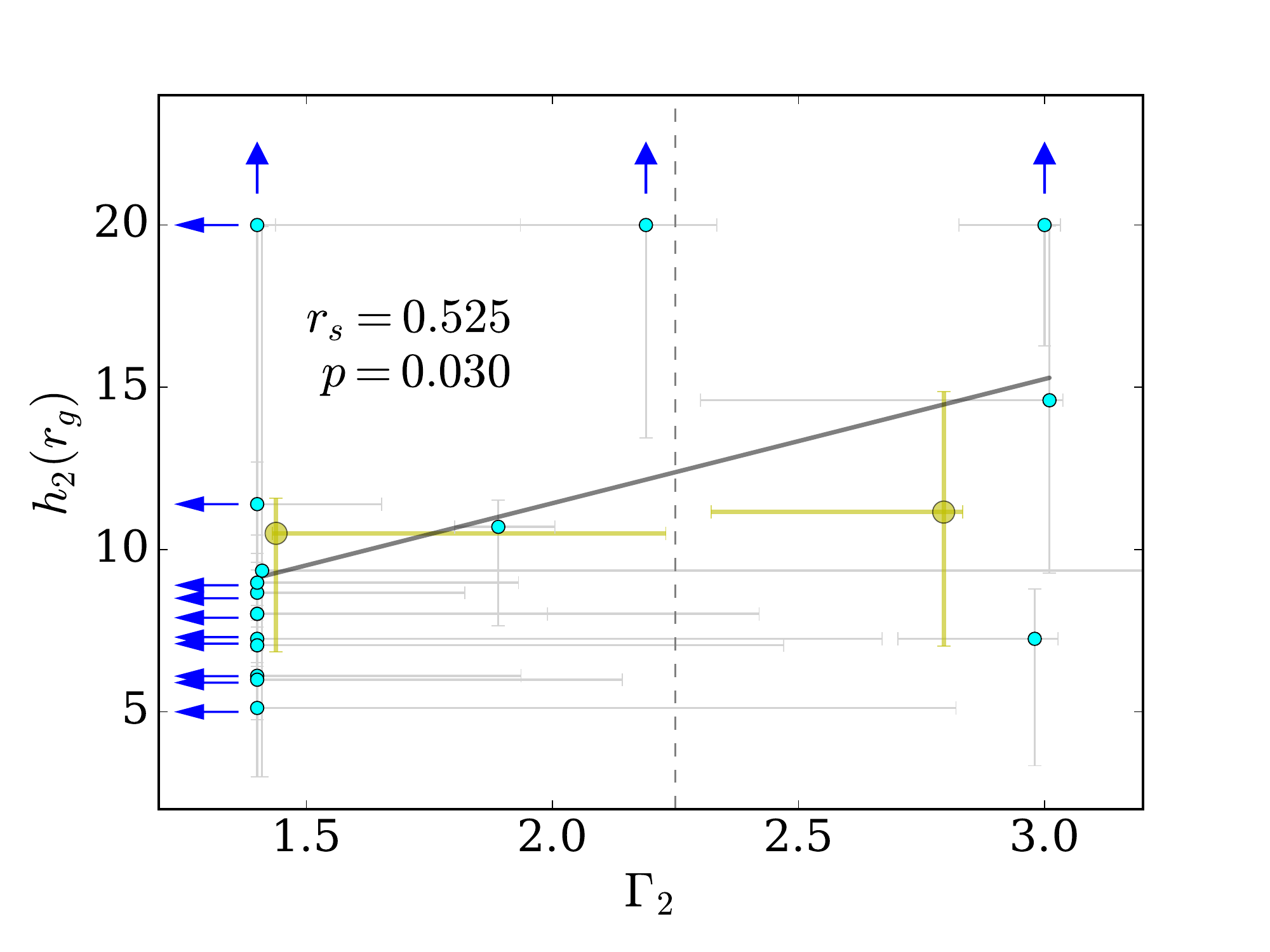}
%plot_IRAS_ECM_h2_g2_VIS.py
\includegraphics[trim={0.10cm 0.2cm 1.5cm 0.2cm},clip,scale=0.30]{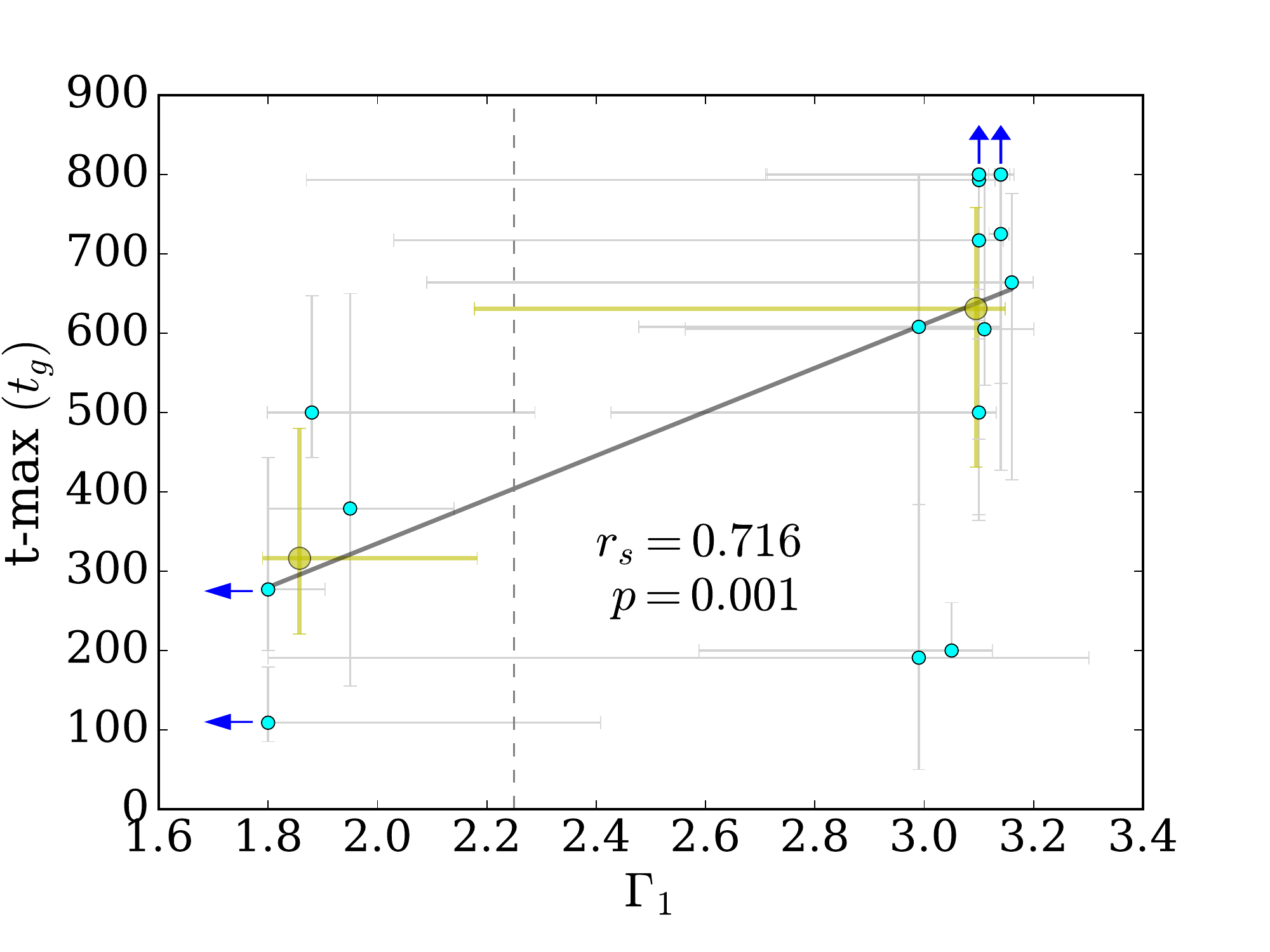}\\
%plot_IRAS_ECM_g1_tmax_VIS.py
\includegraphics[trim={0.1cm 0.2cm 1.5cm 0.2cm},clip,scale=0.30]{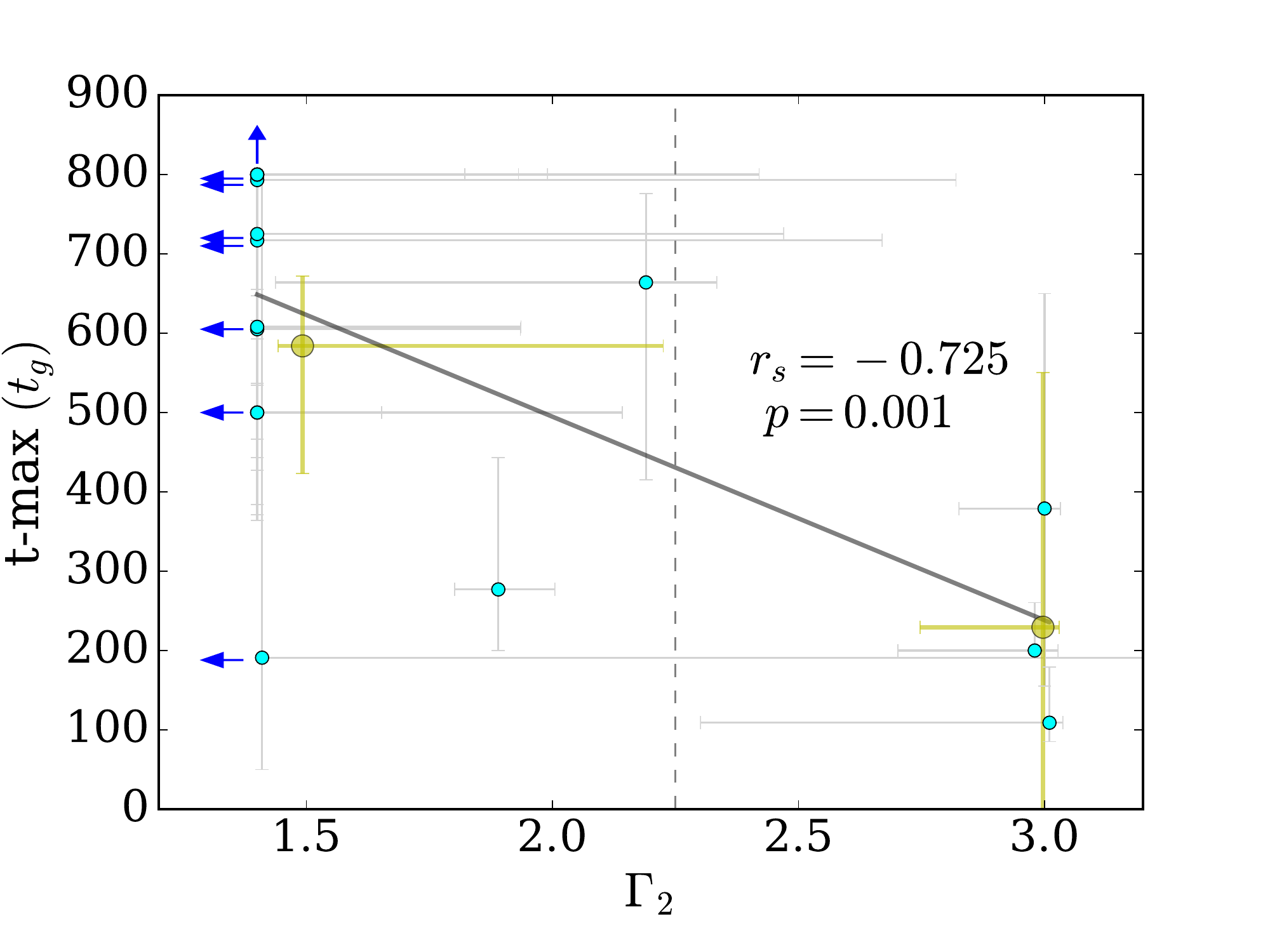}
%plot_IRAS_ECM_g2_tmax_VIS.py
\includegraphics[trim={0.1cm 0.2cm 1.5cm 0.2cm},clip,scale=0.30]{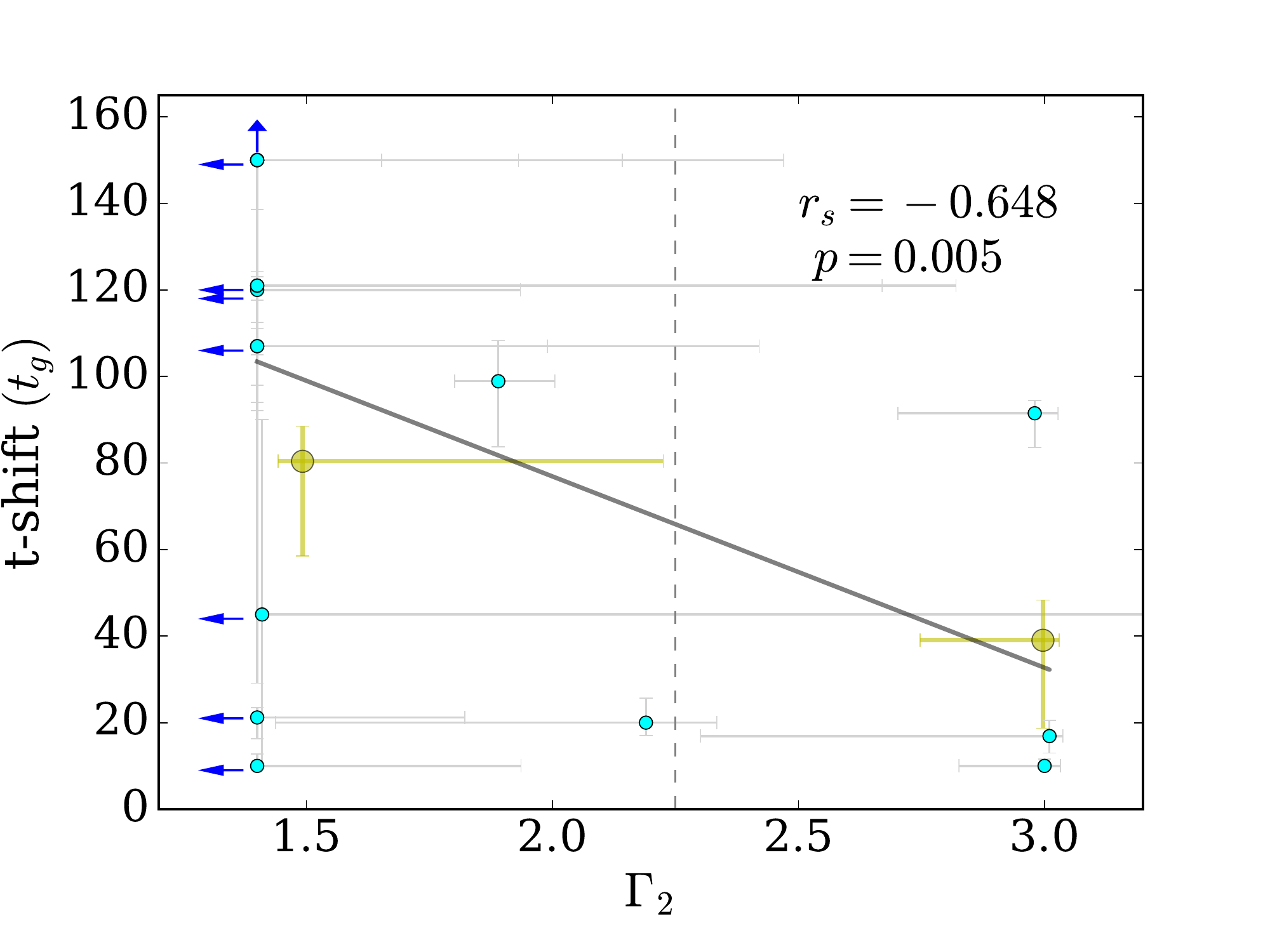}
%plot_IRAS_ECM_g2_tshift_VIS.py
\includegraphics[trim={0.1cm 0.2cm 1.5cm 0.2cm},clip,scale=0.30]{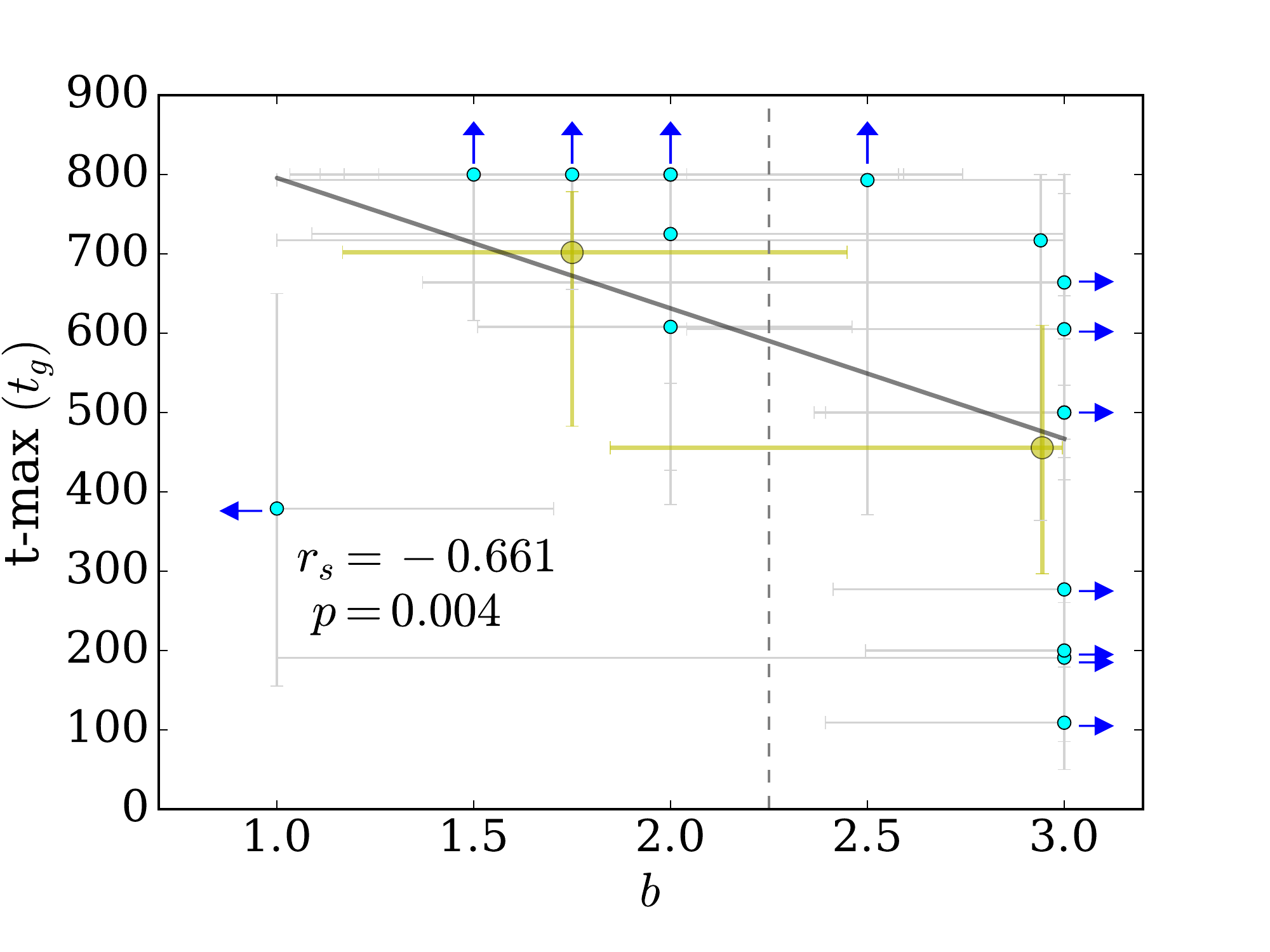}
%plot_IRAS_ECM_b_tmax_VIS.py
\caption[IRAS 13224-3809 correlations]{The IRAS 13224-3809 ECM results showing moderate to strong correlations where the $p$-value \textless{0.05}. Each correlation has been divided into bin sizes as shown by the vertical grey dashed lines and the resultant mean plot (and associated errors) has been shown by the large yellow data points for clarity of each correlation.}
\label{fig:IRAS-SR-results}
\end{figure}

We explore the model for correlations using the Spearman's rank method. For this we use all individual observation table model fits returned from the single mass value phase for each AGN to maximise the available data. We found moderate correlations between $\Gamma_1$ and $t_\text{max}$ and between $\Gamma_2$ and $t_\text{max}$ and $t_\text{shift}$. On the other hand, large errors are still a feature of $t_\text{max}$ therefore we acknowledge that these correlations are not well constrained. Further investigation reveals these correlations are often much stronger in IRAS 13224-3809 than they are in 1H0707-495. The brightness parameters $b$ was strongly anti-correlated with $t_\text{max}$ in IRAS 132224-3809, where no correlation was evident in 1H 0707-495. Note that \cite{2018MNRAS.480.2650C} also suggested that the 1H 0707-495 mass should be fixed at values below $\sim (2-3) \times 10^{6}~M_{\odot}$, otherwise the observed reverberation lags cannot be explained. The time lag amplitude was moderately correlated with the upper photon index $\Gamma_2$ in all cases. The results are presented in Table~\ref{ecm-spearman}, which shows the parameters of interest and the overview of the Spearman's rank correlations $r_s$ and their $p$ values for each source in columns 3 and 4 and for any global correlations in the final column. Note that the covering fraction is also tested from the results reported in HYC22. We found no correlation between the upper source height and the luminosity, however the time lags do correlate moderately with the upper X-ray source height in 1H 0707-495, and a stronger relationship is evident in IRAS 13224-3809. Further consideration of the lags seen $\lesssim 250$~s in the latter source provides a much stronger correlation where $r_s = 0.68$ with $p\text{-value} = 0.005$. The six strongest correlations found for these sources are shown in Figure~\ref{fig:1H0707-SR-results} and \ref{fig:IRAS-SR-results} respectively. The panels within these figures also show where the limits of each parameter were reached as denoted by the blue arrows. 

\begin{table}
\centering
\caption[Extended Corona Model Spearman's rank correlations]{The simultaneously fitted ECM Spearman's rank correlations $r_s$ and the associated $p$ value for each source and all data. Note that the first row is showing the correlations for $h_2 - L_{(2-10~\text{keV})}$ and the final row is the Covering Fraction obtained from HYC22. The last column shows the results when all data from both 1H 0707-495 and IRAS 13224-3809 are used to find the correlation coefficients.} 
\label{ecm-spearman}
\begin{tabular}{ccrcrcrc}
\\ \hline 
{Par1} & {Par2} & \multicolumn{2}{c}{1H 0707-495} & \multicolumn{2}{c}{IRAS 13224-3809} & \multicolumn{2}{c}{1H and IRAS}\\ 
& & {$r_s$} & {$p$} & {$r_s$} & {$p$} & {$r_s$} & {$p$} \\ \hline
$h_2$ & $L$ & 0.052 &  0.853 & -0.127 & 0.628 & -0.200 & 0.272 \\[3pt]
$h_2$ & Lag & 0.312 & 0.257 & 0.556 & 0.020 & 0.418 & 0.017 \\[3pt] 
$h_2$ & $\xi$ & -0.631 & 0.012 & -0.086 & 0.742 & -0.423 & 0.016  \\[3pt]
$h_2$ & $\Gamma_1$ & 0.573 & 0.026 & -0.199 & 0.444 & 0.131 & 0.472 \\[3pt] 
$h_2$ & $\Gamma_2$ & 0.195 & 0.486 & 0.525 & 0.030 & 0.482 & 0.005 \\[3pt]
$\Gamma_1$ & $t_\text{max}$ & 0.216 & 0.440 & 0.716 & 0.001 & 0.591 & 0.001 \\[3pt]
$\Gamma_1$ & $t_\text{shift}$ & 0.327 & 0.235 & 0.303 & 0.237 & 0.381 & 0.031\\[3pt]
$\Gamma_2$ & $t_\text{max}$ & -0.034 & 0.904 & -0.725 & 0.001 & -0.549 & 0.001\\[3pt]
$\Gamma_2$ & $t_\text{shift}$ & -0.587 & 0.021 & -0.648 & 0.005 & -0.594 & 0.00034\\[3pt]
$b$ &  $t_\text{max}$ & -0.345 & 0.125 & -0.661 & 0.004 & -0.411 & 0.019 \\[3pt]
$b$ &  $t_\text{shift}$ & -0.658 & 0.008 & 0.096 & 0.714 & -0.125 & 0.495 \\[3pt]
$b$ & Cvr & -0.603 & 0.017 & 0.214 & 0.408 & -0.154 & 0.402 \\[3pt]
\hline
\end{tabular}
\end{table}

%Whilst $\Gamma_1$ and $\Gamma_2$ correlate with the upper source heights in 1H 0707-495 and IRAS 13224-3809 respectively, 

Both AGN have a moderate relationship for $\Gamma_2$ and $t_\text{shift}$ which can be seen as the strongest global correlation where $r_s = -0.594$ and $p\text{-value} = 0.00034$. The lower source photon index  $\Gamma_1$ also correlates strongly with $t_{\text{max}}$ in IRAS 13224-3809 where $r_s$ = 0.716 and $p\text{-value} = 0.001$ with no counterpart correlation seen in 1H 0707-495. For the upper source $\Gamma_2$ a moderate correlation is seen with the brightness parameter $b$ in 1H 0707-495 and another strong inverse correlation emerged with $t_{\text{max}}$ in IRAS 13224-3809 where $r_s = -0.725$ and $p\text{-value} = 0.001$. Finally, a strong correlation found in IRAS 13224-3809 and not in 1H 0707-495 was $\Delta\Gamma$, ($\Gamma_1 - \Gamma_2$), with $t_\text{max}$ where $r_s = 0.82$ and $p\text{-value} = 5.86 \times 10^{-5}$. It is interesting to note that unlike 1H 0707-495, IRAS 13224-3809 simultaneous fitting does not always lead to $\Gamma_1 > \Gamma_2$. These results suggest that the spectral properties of the corona may be specific to the source. The $\Gamma-L/L_{\text{Edd}}$ and $\Gamma-h_2$ relationships for these AGN are also presented in Figure~\ref{fig:gamma_EddFrac}. It can be seen that lower and higher limits of $\Gamma$ and $h_2$ often reached their limits and these parameters may be wider as indicated by the blue arrows in each panel. 

In fact, the photon index $\Gamma_1$ tends to increase sharply with increasing luminosity above $\sim2.5 \times 10^{42}$ erg cm s$^{-1}$ for both sources. A similar trend in AGN data against the luminosity ratio $L_x / L_\text{Edd}$ has been reported by \cite{2015MNRAS.447.1692Y} and interpreted as a two-phase accretion flow model, although more data would be required to investigate this scenario. For 1H 0707-495, the correlation is flat at lower luminosity then a mild positive correlation kicks in when the luminosity is $10^{-1.7}\lesssim L/L_{\text{Edd}}\lesssim 10^{-1.5}$. Contrarily, for IRAS 13224-3809, we see the progressively flat profile of $\Gamma$ instead, and their $\Gamma_1$ and $\Gamma_2$ values are extreme and more separated. The behaviour of $\Gamma$ may suggest that contrasting mechanisms are driving the variability in these AGN (within the model constraints). 

\section{Discussion}

The initial fitting of the ECM to the various flux groups achieved reasonable fits and model descriptions, however these were not always well constrained and error ranges often tended to the model maximum and minimum allowed values. The model can provide a good statistical fit to the observed data when the mass is employed as a free parameter and is capable of finding the upper X-ray source heights between 3 -- 20\ $r_g$. Fixing the mass at the single best value obtained from the simultaneous fits also achieved similar results and whilst the black hole mass of 1H 0707-495 dropped from 6.22 (obtained by the independent emcee fit) to $6.04 \pm 0.01 \log(M/M_{\odot})$, both model fits remain statistically reasonable. The black hole mass for IRAS 13224-3809 was estimated at $6.30 \pm 0.01 \log(M/M_{\odot})$ and remained reasonably consistent with the independent measurement. These obtained mass values contained very small errors and they fall within the limits of the variable mass values from the model fits presented in Table \ref{1H_TM_ECM_fits}.

\cite{2020MNRAS.498.3184C} fix the mass of IRAS 13224-3809 at $\log (M/M_{\odot})=6.30$ and their model looks at the reverberation from a single point source and predicts only the negative reverberation lags, whilst the positive lags at low frequency are modelled separately using the \texttt{KYNXILREV} phenomenological power law. Our model is mass dependent and considers both soft and hard lag mechanisms simultaneously. 
Nevertheless, we note that a limiting factor may be that we have fixed the lower source height at $2\ r_g$; an assumption required to ease the heavy computations required to create the predicted time lags and hence we are estimating the \textit{extent} of the corona rather than its true size that may be obtained by employing $h_1$ as a free parameter. This would be extremely intensive with the current model and further development should maintain its complexity whilst enhancing model computational performance.

Despite this, some meaningful relations among the ECM parameters can still be inferred. For 1H 0707-495, we find significant correlation ($p<0.05$) between $h_2$ and $\Gamma_1$, and significant anti-correlation between $h_2$ and $\xi$. This means that when the 1H 0707-495 corona extends upwards (larger $h_2$), the coronal emission seems to be softer (larger $\Gamma$) while producing less overall ionisation on the disc (smaller $\xi$). This is expected since the vertically extended corona should produce less intense illumination pattern on the inner disc \citep[e.g.][]{2012MNRAS.424.1284W}. The tendency of softer corona with increasing its vertical extent is also found in IRAS 13224-3809. This is in line with \cite{alston_dynamic_2020} and \cite{2022ApJ...934..166C} where the lamp-post geometry was used and the photon index of the continuum was found to increase with the source height. In fact, \cite{2022ApJ...934..166C} also found that during the source height increases, the disc itself generates more high-frequency variability, suggesting that the inner disc becomes more active. Here, we find the significant anticorrelation between $\Gamma_2$ and both $t_{\text{max}}$ and $t_{\text{shift}}$ ($r_{s}=-0.724$ and $-0.648$, respectively). Therefore, when the extending corona becomes softer, the source response as well as the signal propagation occurs faster. This may agree with \cite{2022ApJ...934..166C} that as the source height increases the innermost region is more active so that the corona requires shorter time in response.

Nevertheless, in the ECM environment, the time lags in IRAS 13224-3809 correlate stronger with $h_2$ than those found in 1H 0707-495. Consideration of the lags seen under 250 s provides a much stronger correlation where $r_s = 0.68$ with $p$-value = 0.005, hence there are hints that much longer time lags will still follow this correlation albeit with a shallower slope. Furthermore, \cite{Chainakun2019} pointed out that in order reproduce the time lags, their spherical corona model required higher coronal temperatures for a lower optical depth, $\tau$, supporting the $\tau$ and coronal temperature anti-correlation argument discussed by \cite{Tortosa2018}. Recently, \cite{2022MNRAS.513..648C} presented a new method of predicting the black hole mass of X-ray reverberating AGN using artificial neural networks and pointed out that the inconsistency of the lag-mass relationship may be due to the lag amplitudes being more strongly affected by other geometric effects that are not related to the mass of the black hole. Also, there is a significantly larger number of parameters in the ECM than in the lamp-post model. Perhaps, this might explain the observed anti-correlation between the lags and the mass seen in IRAS 13224-3809 under the extended corona environment when using individual observations (Fig.~\ref{fig:TM-vari-M-lag}), that the lag-mass relation is also modulated by other coronal parameters in the model.

In the lamp-post geometry, the X-ray source heights in IRAS 13224-3809 have been reported to correlate positively with the luminosity in the 2 -- 10 keV energy range \citep{alston_dynamic_2020,2020MNRAS.498.3184C,2022ApJ...934..166C}. However, there was no evidence of this relationship in this study. Whilst this is consistent with the findings of \cite{2020A&A...641A..89S}, it is in contrast to their findings that the source is compact within a few gravitational radii (when modelled using the relativistically smeared reflection from the accretion disc). Variations of the inner regions and accretion flow have been explained via transitions between the jet emitting disc (JED) and standard accretion disc (SAD) framework \citep{2022A&A...659A.194M} and the framework of coupled hot accretion and jet model for GBHBs and AGN \citep{2015MNRAS.447.1692Y}, suggesting that a two-phase accretion flow could be driving the observed variability.

\onecolumn
% TRIM={A,B,C,D} = {Left, Bottom, Right, Top}
% eg:  [trim={0.7cm 0.2cm 0.5cm 0.2cm},clip,scale=0.38]{images/1H0707_tm_comb.pdf}

% TRIM={A,B,C,D} = {Left, Bottom, Right, Top}
% eg:  [trim={0.0cm 0 1.5cm 0.8cm},clip,scale=0.6]{images/1H0707_tm_comb.pdf}

\twocolumn

%\onecolumn
\begin{figure}
\centering
\includegraphics[trim={0.2cm 0.2cm 0.4cm 0.2cm},clip,scale=0.367]{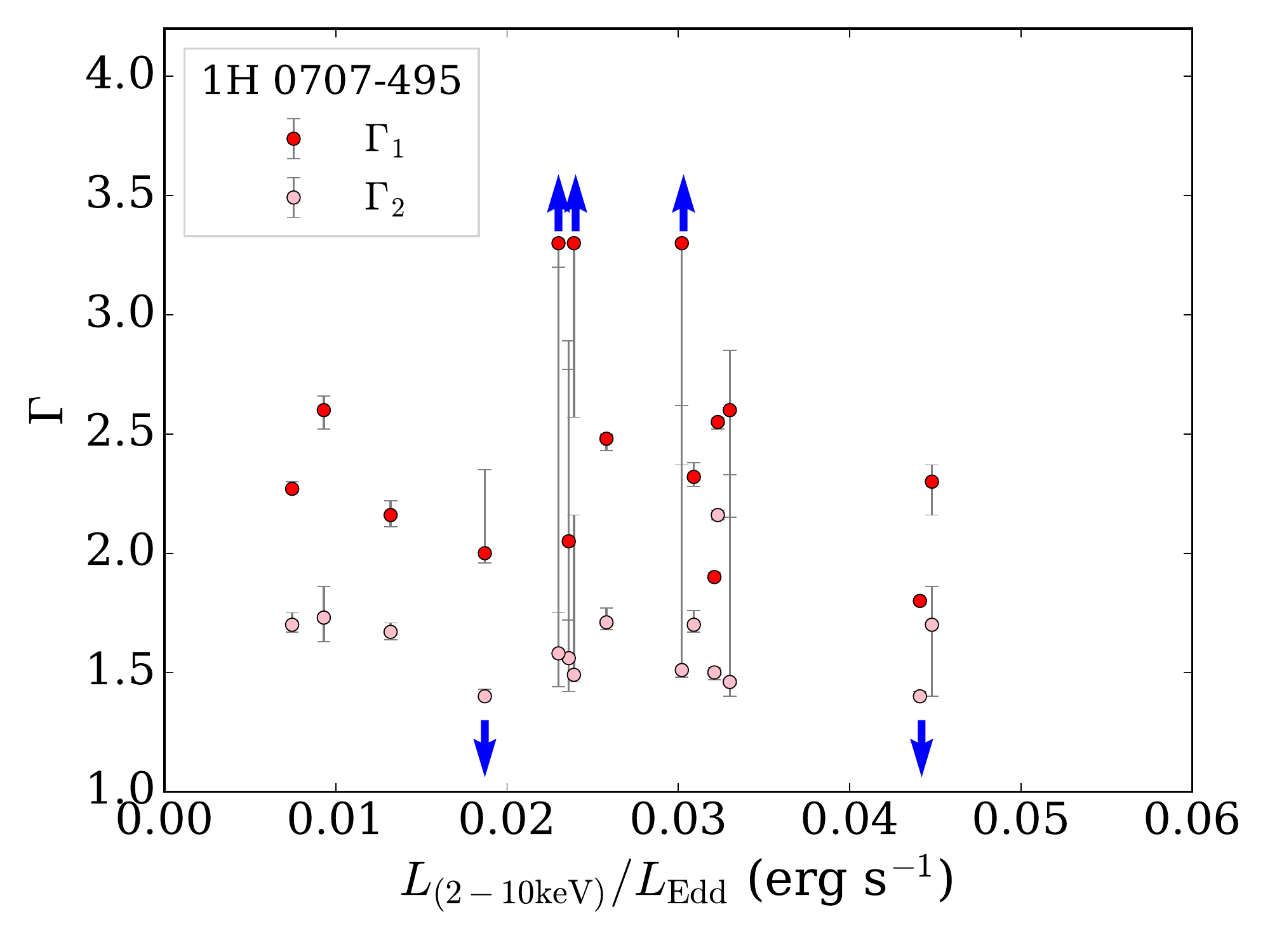}
\includegraphics[trim={0.1cm 0.1cm 0.4cm 0.2cm},clip,scale=0.367]{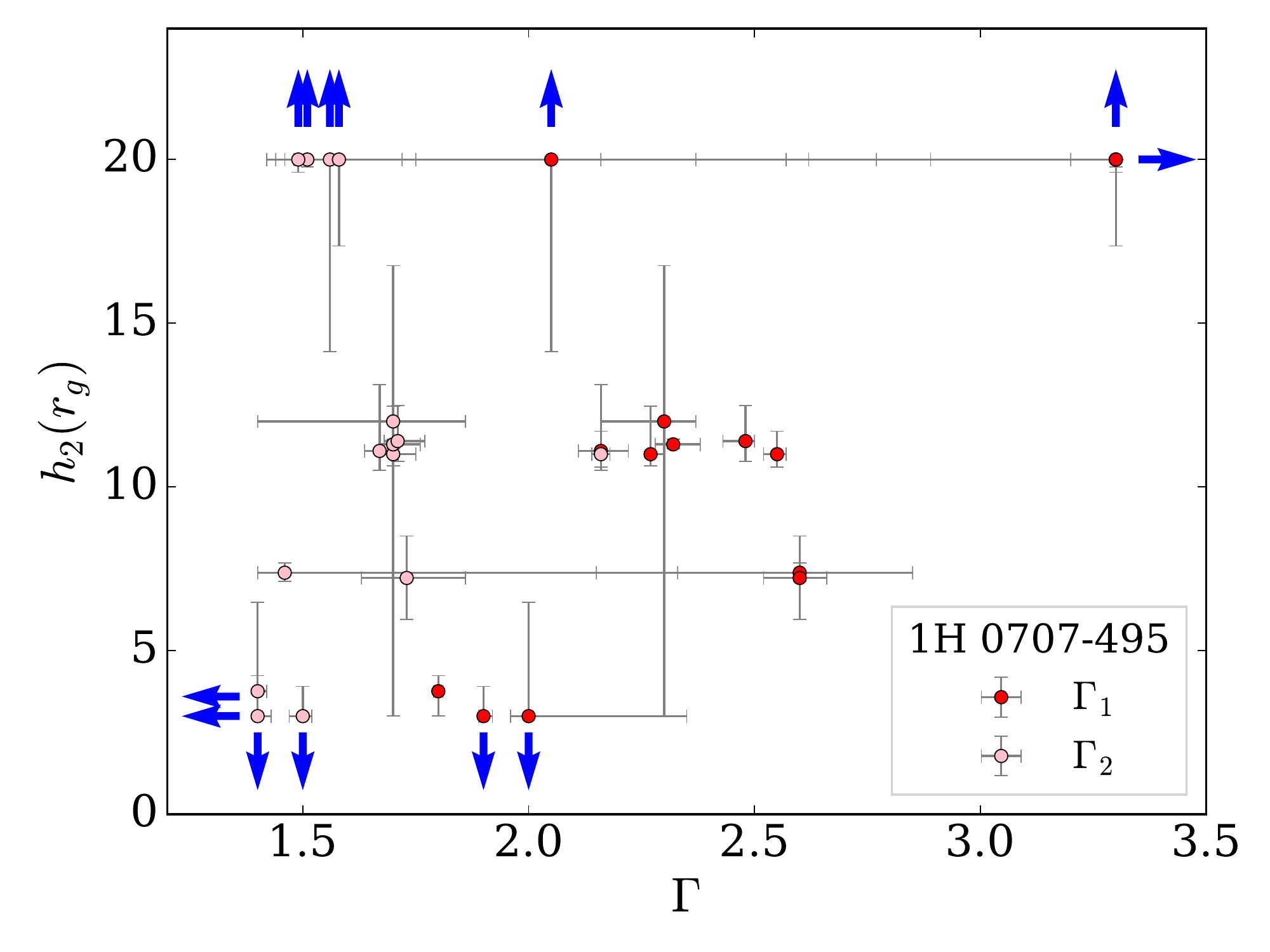}\\
\includegraphics[trim={0.2cm 0.2cm 0.4cm 0.2cm},clip,scale=0.367]{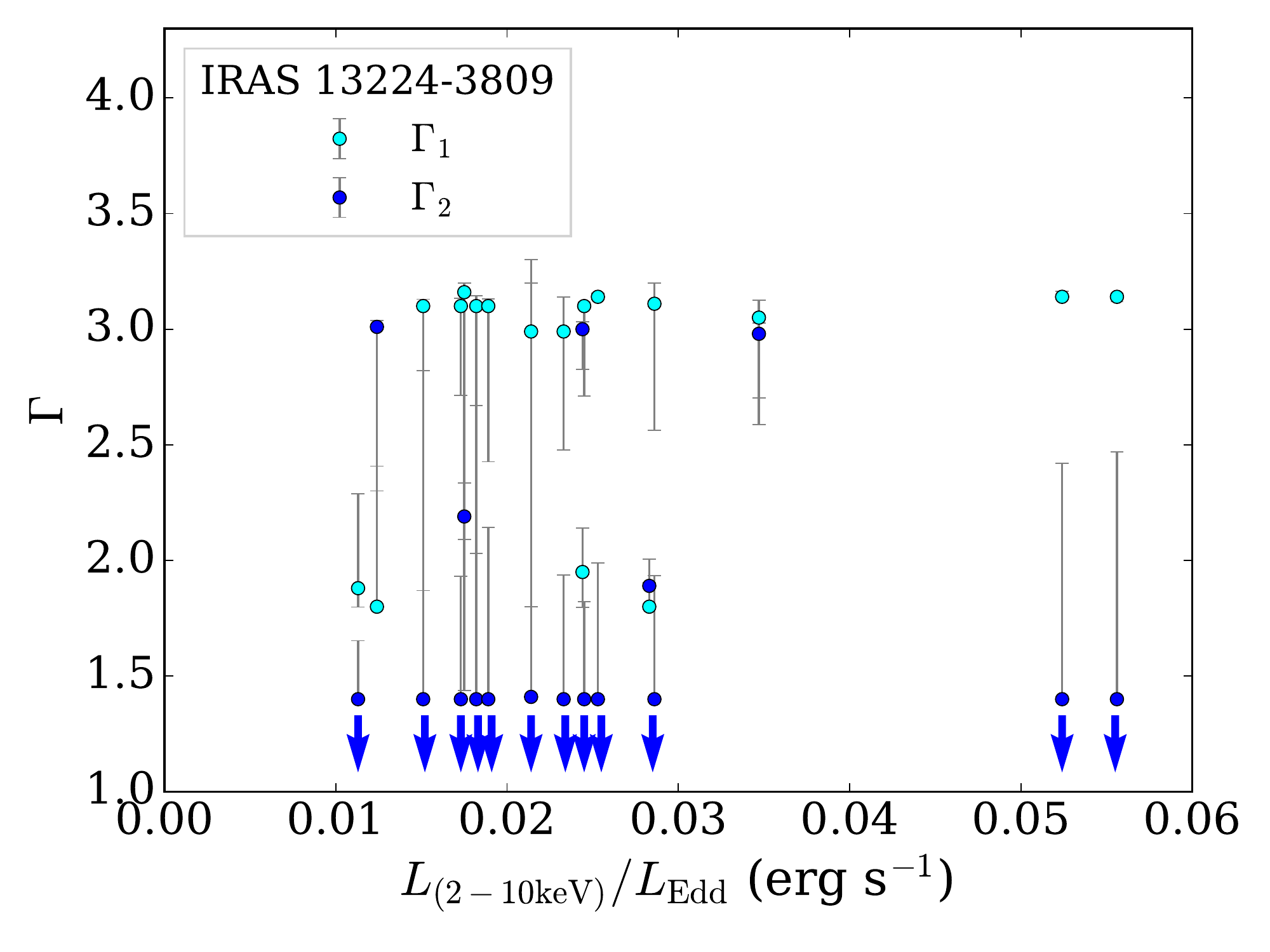}
\includegraphics[trim={0.1cm 0.1cm 0.4cm 0.2cm},clip,scale=0.367]{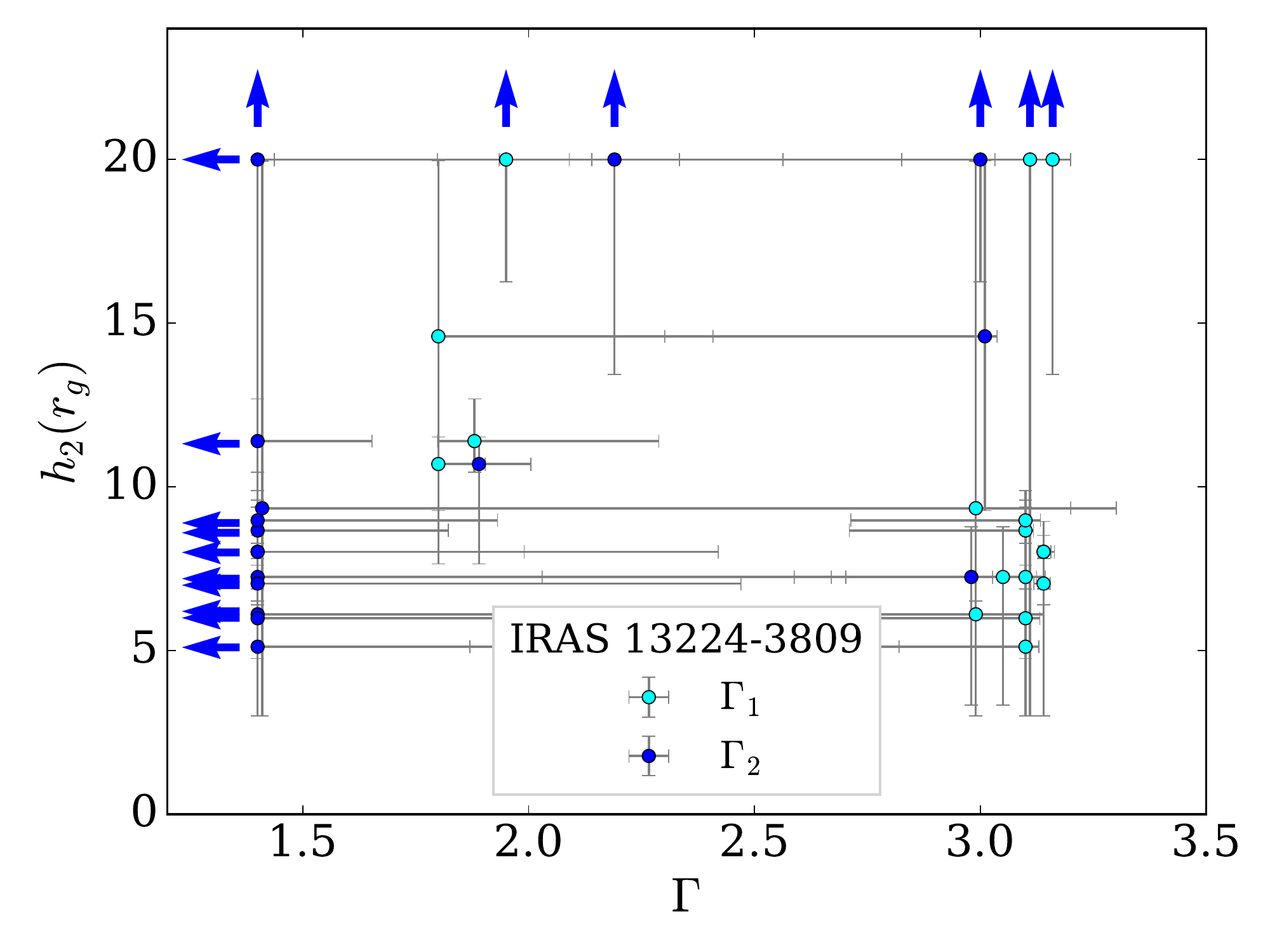}
\caption[1H 0707-495 and IRAS 13224-3809 $\Gamma$ and Eddington fraction]{The evolution of $\Gamma$ as a function of the Eddington fraction and the source height for 1H 0707-495 \emph{(top panels)} and also for IRAS 13224-3809 \emph{(bottom panels)}. The $\Gamma_1$ and $\Gamma_2$ values IRAS 13224-3809 are more separated and occupied at more extreme values than those of 1H 0707-495, especially when $5\ r_g \lesssim h_2 \lesssim 10\ r_g$.} \label{fig:gamma_EddFrac}
\end{figure}

%We find the source could extended beyond $20\ r_g$ in some cases, where the only moderate correlation was found in the the upper X-ray source height and the lower X-ray source photon index. Whilst our estimation of the time lags follow the same profile as seen in many of these studies, they originated from different groups, estimated down to the lowest frequencies appropriate to the duration of the observations in each group.

For 1H 0707-495, the profile of $\Gamma$ is flat at lower luminosity before showing a mild positive correlation at $10^{-1.7}\lesssim L/L_{\text{Edd}}\lesssim 10^{-1.5}$ (Figure~\ref{fig:gamma_EddFrac}). Although this is a small fraction of the Bolometric luminosity, similar findings have been reported by the two-phase accretion Type II Luminous Hot Accretion Flow (LHAF) and disc-corona regime, after which the evolutionary turnover point is reached when the accretion rate changes. Furthermore, the evolutionary behaviour of $\Gamma_1$ and $\Gamma_2$ is very similar when tracked against the Eddington fraction. Although there is not a multitude of individual AGN categorised in the two-phase accretion scenario due to the requirement for multiple long observations, a theoretical explanation of this behaviour was found in AGN and reported by \cite{2003MNRAS.342..355Z} who suggested that the emission of a cold medium irradiating hot plasma could be due to the cooling via reprocessed emission of the hot plasma causing the observed X-ray spectrum to soften, leading to a higher value of $\Gamma$. 

Contrarily, the $\Gamma$ profile for IRAS 13224-3809 is relatively flat, which is comparable to the $\Gamma - L_{\text{Bol}}/L_{\text{Edd}}$ behaviour of the jet-phase regime of the JED-SAD framework where the $\Gamma$ is determined by the energy distribution of power law electrons in the jet and roughly constant for sources with different luminosity's, neither source follows the expected $\Gamma \approx 2.1$. Furthermore, for IRAS 13224-3809, the current model values of $\Gamma_1$ are extreme although not uncommon \citep[see e.g.,][]{2014MNRAS.443.2746W} and $\Gamma_2$ may drop well below 1.4 approaching lower extreme values. There appears to be no relationship with either source $\Gamma$ and the accretion rate in IRAS 13224-3809. Similar $\Gamma$ behaviour was reported from \textit{RXTE} observations of MCG-6-30-15 where the spectral index increased with luminosity, reaching a final upper value after which it remained roughly static as the luminosity continued to increase \citep{MCHARDY1999}, possibly due to a harder spectral component that does not change within the observation duration as the continuum component steepens the spectra. 

A steeper spectrum is expected for increasing luminosity and this is evident in the spectral model for IRAS 13224-3809 as reported in HYC22. The ECM scenario, however, shows very limited spectral index variations for its flux variations. This has been previously reported for NGC 5548 \citep{2009MNRAS.399.1597S} and PG 0804+761 \citep{2003MNRAS.344..993P}. Furthermore, \cite{2012A&A...537A..87C} found no significant spectral variability in NGC 4151 and NGC 2110, but reported the significant flux variations that indicated the intrinsic variability of the central source in NGC 4388, NGC 4945 and IC 4329. This raises the question of different spectral states in AGN. However, 1H 0707-495 and IRAS 13224-3809 are both classified Narrow-line Seyfert 1 galaxies where the 2--10\,keV flux variations are almost always associated with spectral variations \citep{2002ApJ...573...92P}. The modelling applied thus far (spectral and ECM models) suggests that different mechanisms may be contributing to the variability that may be explained by the differences in the geometry of each source. 

Note that variations in the optical depth $\tau$ of the corona cause spectral changes in the Comptonised emission leading to a steeper spectrum and the temperature decreases with increasing $\tau$ \citep[e.g.][]{1997ApJ...476..620H}, therefore the geometry of 1H 0707-495 should consist of a smooth corona where the intrinsic flux varies in unison with $\Gamma_1$ and $\Gamma_2$. IRAS 13224-3809 however, may contain a slightly different geometry since the $\Gamma_1$ and $\Gamma_2$ values are more clearly separated, generally remaining relatively constant and occupied at the the extreme values of the model especially when $5\ r_g \lesssim h_2 \lesssim 10\ r_g$ (Figure~\ref{fig:gamma_EddFrac}). This suggests that IRAS 13224-3809 corona is more patchy (i.e. having clear, distinct spectral zones of spectral $\Gamma_1$ and $\Gamma_2$ that may be related to two distinct zones of corresponding $\tau$ and coronal temperature), especially at locations where the corona $\lesssim 10\ r_g$.

Global coronal correlations across AGN samples have remained elusive and/or confusing \citep[see e.g.,][]{Sarma2015,2021MNRAS.tmp.1759H,Kamraj2022} the latter authors cautioning the use of coronal parameters and the Eddington fraction relations to infer properties of black hole systems. In contrast, this work finds that strong relations are evident in the grouped data and when drilling down to individual observations. These differences are possibly due to the choice of spectral models and parameters explored. 

Many of the $\chi^2$ values fell below unity, suggesting that the model may be over-fitted due to too many free parameters. This suggests that deeper model assumptions leading to the freezing of at least one more parameter may also be appropriate for future modelling. Some parameter ranges may need to be slightly wider to accommodate better statistics (whilst remaining feasible), especially for $h_2$, $\Gamma_1$ and $\Gamma_2$. On the other hand, the large errors seen in $t_{\text{max}}$ and $t_{\text{shift}}$ will be difficult to constrain with current observations and the large errors associated with the low frequency fluctuation lags. More AGN need to be explored by this model to test its capabilities against a wider sample with higher quality time lags versus frequency allowing a better statistical comparison of the data and models. 

\section{Conclusions}

This work has explored the extended corona scenario using two X-ray point sources. The lower source may represent the base of a jet-like structure or the lower regions of a compact corona with the upper source representing the extended region where the flares are detected due to the periodic vertical collimation of a jet structure or outer region of the corona. The model is capable of fitting the lag-frequency of 1H 0707-495 and IRAS 13224-3809, where their black hole mass can be constrained to $\log (M/M_{\odot})=6.04\pm0.01$ and $6.30\pm0.01$, respectively. 

This work also supports the advantages of exploring corona correlations using individual observations and also suggests that the physics and geometry of the corona may diverge between different sources. For the first time, we have gained an insight to the behaviour of $\Gamma$ for each of the X-ray sources located above the black hole axis through the use of reverberation mapping. We find the tendency of softer corona with increasing its vertical extent in both AGN. This produces a much less-ionised disc which is evident in 1H 0707-495. On the other hand, for IRAS 13224-3809 we also find the hint of increasing $\Gamma_2$ with decreasing $t_{\text{max}}$ and $t_{\text{shift}}$, suggesting that shorter propagating fluctuations and faster coronal response may be evident when the corona extends vertically upwards. The intrinsic flux of IRAS 13224-3809 show less variability with $\Gamma$ and its corona may be more patchy in a sense that contains clearer separate spectral zones of distinct $\Gamma_1$ and $\Gamma_2$ that may also link to two clearer distinct zones of coronal temperature than that of 1H 0707-495.

\section*{Acknowledgements}

The calculations in this work were carried out using the high performance computer \texttt{BlueCrystal} of the Advanced Computing Research Centre, University of Bristol, UK. SH thanks the STFC for funding and the Bristol, Cardiff \& Swansea CDT Team for support. PC thanks funding support from (i) Suranaree University of Technology (SUT), (ii) Thailand Science Research and Innovation (TSRI), and (iii) National Science Research and Innovation Fund (NSRF), project no. 160355. This research has made use of ISIS functions (ISISscripts) provided by
ECAP/Remeis observatory and MIT (http://www.sternwarte.uni-erlangen.de/isis/). We wish to thank the anonymous reviewer for comments and suggestions which has improved the quality of this manuscript.

\section*{Data availability}
The original data underlying this article were obtained from the \emph{XMM-Newton} Observatory (\url{http://nxsa.esac.esa.int}). The derived data generated in this research can be downloaded via \url{http://www.star.bris.ac.uk/steff/hancock.html}. 
\bibliographystyle{mnras}
\bibliography{ecm}

%\clearpage

\onecolumn
\appendix

\section{Table model ECM results}
\begin{table}
\centering
\caption[ECM Table Model fits]{The Table Model results for 1H 0707-495 and IRAS 13224-3809 data to $90\%$ confidence.  Note that zero errors are a result of rounding to 3 sf and also hitting upper or lower model parameter space limits.} 
\label{1H_TM_ECM_fits}
%\begin{adjustbox}{width=\textwidth}
\begin{tabular}{ccccccccccc}
\hline \\%[1pt] 
{Source} & {Obs Id} & {$\chi^2$}  & {$\log (M/M_\odot$)} & {$h_2 (r_g)$} & {$\Gamma_1$}  & {$\Gamma_2$} & {log$\xi (\text{erg cm s}^{-1})$} & {$b$} & {$t_\text{max} (t_g)$} & {$t_\text{shift} (t_g)$}\\[1pt] 
\hline
\\[2pt] 
1H & Combined & 0.50 & $6.11^{+0.01}_{-0.03}$ & $3.49^{+0.35}_{-0.36}$ & $2.40^{+0.00}_{-0.00}$ & $1.94^{+0.01}_{-0.01}$ & $0.01^{+0.11}_{-0.01}$ & $3.00^{+0.00}_{-0.04}$ & $370.87^{+13.13}_{-13.27}$ & $17.06^{+0.20}_{-0.20}$   \\[3pt] 
& Hi-flux  & 0.07 & $6.21^{+0.00}_{-0.17}$ & $10.04^{+0.71}_{-0.70}$ & $2.69^{+0.11}_{-0.22}$ & $1.93^{+0.43}_{-0.03}$ & $2.01^{+0.99}_{-2.00}$ & $3.00^{+0.18}_{-0.28}$ & $308.29^{+18.39}_{-27.68}$ & $15.91^{+4.96}_{-2.49}$ \\[3pt]
%& Hi-counts & 0.00 & $6.37^{+0.06}_{-0.13}$ & $15.15^{+4.85}_{-0.00}$ & $2.66^{+0.14}_{-0.06}$ & $1.90^{+0.15}_{-0.00}$ & $2.44^{+0.56}_{-2.44}$ & $2.52^{+0.17}_{-0.00}$ & $234.42^{+0.04}_{-2.58}$ & $21.04^{+0.00}_{-0.08}$ \\ [3pt]
& Med-flux & 1.11 & $5.31^{+0.19}_{-0.03}$ & $5.03^{+14.97}_{-0.87}$ & $2.57^{+0.24}_{-0.05}$ & $2.47^{+0.14}_{-0.50}$ & $0.04^{+2.95}_{-0.05}$ & $1.77^{+1.23}_{-0.77}$ & $667.13^{+32.87}_{-567.13}$ & $71.68^{+8.52}_{-46.25}$ \\[3pt]
& Lo-flux & 1.57 & $6.21^{+0.26}_{-0.39}$ & $3.00^{+2.99}_{-0.00}$ & $3.10^{+0.00}_{-0.03}$ & $1.90^{+0.49}_{-0.00}$ & $2.99^{+0.00}_{-0.99}$	& $1.94^{+1.06}_{-0.94}$ & $615.53.^{+34.31}_{-515.53}$ & $59.99^{+46.70}_{-23.68}$ \\[3pt]
%& Lo-counts & 0.46 & $5.89^{+0.18}_{-0.11}$ & $3.00^{+1.70}_{-0.00}$ & $2.41^{+0.18}_{-0.00}$ & $2.25^{+0.19}_{-0.05}$ & $0.02^{+2.07}_{-0.02}$ & $2.13^{+0.87}_{-0.68}$ & $616.61^{+83.39}_{-101.36}$ & $45.16^{+13.32}_{-1.97}$ \\[3pt]
& 110890201 &	1.74 & $5.84^{+0.39}_{-0.22}$ & $19.30^{+0.67}_{-16.30}$ & $2.13^{+0.29}_{-0.13}$ & $1.70^{+0.80}_{-0.00}$ & $0.01^{+2.99}_{-0.01}$ & $1.87^{+0.48}_{-0.87}$ & $700.00^{+49.80}_{-297.00}$ & $43.70^{+56.30}_{-33.70}$	 \\[3pt]
& 148010301 &	0.92 & $6.51^{+0.05}_{-0.01}$ & $16.00^{+4.00}_{-11.20}$ & $2.70^{+0.30}_{-0.53}$ & $1.70^{+0.80}_{-0.00}$ & $2.49^{+0.51}_{-2.49}$ & $3.00^{+0.00}_{-0.48}$ & $350.00^{+345.00}_{-217.00}$ & $18.50^{+16.20}_{-8.47}$	 \\[3pt]
& 506200201 &	0.12 & $6.54^{+0.10}_{-0.54}$ & $3.58^{+7.67}_{-0.58}$ & $2.53^{+0.47}_{-0.49}$ & $	1.70^{+0.80}_{-0.00}$ & $2.63^{+0.38}_{-2.63}$ & $1.95^{+1.05}_{-0.95}$ & $416.00^{+280.00}_{-316.00}$ & $50.00^{+35.10}_{-14.30}$	 \\[3pt]
& 506200301 &	0.00 & $6.15^{+0.01}_{-0.01}$ & $18.20^{+1.81}_{-0.56}$ & $2.37^{+0.07}_{-0.07}$ & $1.72^{+0.09}_{-0.02}$ & $0.01^{+2.99}_{-0.01}$ & $2.76^{+0.09}_{-0.09}$ & $518.00^{+22.70}_{-28.600}$ & $34.10^{+0.70}_{-0.70}$ \\[3pt]
& 506200401 &	0.81 & $6.49^{+0.01}_{-0.02}$ & $13.60^{+0.24}_{-0.24}$ & $2.80^{+0.13}_{-0.02}$ & $1.70^{+0.02}_{-0.00}$ & $0.03^{+2.12}_{-0.03}$ & $2.87^{+0.13}_{-0.15}$ & $440.00^{+48.30}_{-36.00}$ & $20.50^{+0.67}_{-1.10}$	 \\[3pt]
& 506200501 & 2.75 & $6.42^{+0.02}_{-0.03}$ & $17.50^{+2.55}_{-0.73}$ & $2.84^{+0.12}_{-0.12}$ & $1.70^{+0.13}_{-0.00}$ & $0.70^{+2.30}_{-0.70}$ & $2.63^{+0.21}_{-0.20}$ & $434.00^{+36.40}_{-105.00}$ & $21.10^{+1.38}_{-2.13}$ \\[3pt]
& 511580101 &	0.81 & $6.48^{+0.07}_{-0.07}$ & $4.07^{+1.50}_{-1.07}$ & $2.99^{+0.01}_{-0.36}$ & $2.70^{+0.00}_{-1.00}$ & $2.92^{+0.08}_{-0.61}$ & $3.00^{+0.00}_{-1.63}$ & $400.00^{+99.60}_{-250.00}$ & $120.00^{+0.00}_{-11.70}$	 \\[3pt]
& 511580201 &	0.38 & $6.03^{+0.26}_{-0.07}$ & $4.14^{+15.90}_{-1.14}$ & $2.85^{+0.15}_{-0.65}$ & $1.70^{+0.80}_{-0.00}$ & $0.01^{+1.14}_{-0.01}$ & $1.00^{+1.29}_{-0.00}$ & $489.00^{+46.00}_{-286.00}$ & $110.00^{+10.00}_{-50.00}$ \\[3pt]
& 511580301 &	0.00 & $6.45^{+0.06}_{-0.11}$ & $16.50^{+3.05}_{-5.52}$ & $2.59^{+0.00}_{-0.07}$ & $1.73^{+0.14}_{-0.03}$ & $0.00^{+3.00}_{-0.00}$ & $2.37^{+0.24}_{-0.39}$ & $223.00^{+2.07}_{-23.40}$ & $1.20^{+0.14}_{-2.09}$	\\[3pt]
& 511580401 &	1.18 & $6.63^{+0.04}_{-0.95}$ & $20.00^{+0.02}_{-1.72}$ & $2.30^{+0.70}_{-0.20}$ & $1.70^{+0.79}_{-0.00}$ & $2.99^{+0.01}_{-1.61}$ & $2.17^{+0.52}_{-0.44}$ & $100.00^{+43.00}_{-0.00}$ & $25.02^{+1.39}_{-3.52}$	 \\[3pt]
& 554710801 &	0.00 & $5.94^{+0.00}_{-0.21}$ & $17.00^{+3.00}_{+0.32}$ & $2.60^{+0.70}_{-0.53}$ & $2.10^{+0.02}_{-0.40}$ & $0.00^{+3.00}_{-0.00}$ & $3.00^{+0.00}_{-0.32}$ & $649.00^{+100.00}_{-131.00}$ & $29.2^{+7.12}_{-9.21}$ \\[3pt]
& 653510301 &	0.01 & $6.55^{+0.02}_{-0.10}$ & $13.30^{+3.34}_{-0.00}$ & $2.88^{+0.02}_{-0.00}$ & $2.26^{+0.18}_{-0.09}$ & $0.04^{+2.96}_{-0.04}$ & $3.00^{+0.00}_{-0.05}$ & $182.00^{+0.00}_{-0.00}$ & $19.20^{+4.38}_{-0.00}$ \\[3pt]
& 653510401 &	0.00 & $6.71^{+0.01}_{-0.00}$ & $17.30^{+2.70}_{-3.67}$ & $2.82^{+0.35}_{-0.18}$ & $1.85^{+0.30}_{-0.15}$ & $0.04^{+2.96}_{-0.04}$ & $2.71^{+0.20}_{-0.03}$ & $105.00^{+41.10}_{-5.37}$ & $17.30^{+0.00}_{-0.00}$	 \\[3pt]
& 653510501 &	0.84 & $6.48^{+0.02}_{-0.11}$ & $3.32^{+14.50}_{-0.32}$ & $2.94^{+0.13}_{-0.33}$ & $1.83^{+0.38}_{-0.13}$ & $3.00^{+0.00}_{-0.99}$ & $3.00^{+0.00}_{-1.10}$ & $500.00^{+129.00}_{-231.00}$ & $10.00^{+4.32}_{-0.00}$	\\[3pt]
& 653510601 &	0.00 & $6.60^{+0.04}_{-0.00}$ & $19.60^{+0.35}_{-9.64}$ & $2.90^{+0.40}_{-0.22}$ & $2.45^{+0.02}_{-0.00}$ & $1.93^{+1.07}_{-1.93}$ & $3.00^{+3.00}_{-0.73}$ & $268.00^{+0.00}_{-0.55}$ & $25.80^{+0.02}_{-0.02}$ \\[3pt] 
\\[2pt] 
IRAS & Combined & 0.00 & $6.66^{+0.01}_{-0.02}$ & $12.73^{+0.00}_{-0.00}$ & $2.62^{+0.24}_{-0.00}$ & $1.70^{+0.02}_{-0.00}$ & $0.99^{+2.00}_{-0.99}$ & $2.48^{+0.52}_{-0.45}$ & $305.23^{+40.93}_{-30.48}$ & $19.45^{+1.04}_{-0.00}$   \\[3pt] 
& Hi-flux  & 4.27 & $6.36^{+0.00}_{-0.00}$ & $4.88^{+1.21}_{-1.88}$ & $3.10^{+0.00}_{-0.72}$ & $1.70^{+0.73}_{-0.00}$ & $3.00^{+0.00}_{-3.00}$ & $2.67^{+0.33}_{-0.97}$ & $956.33^{+543.66}_{-27.41}$ & $150.00^{+0.00}_{-0.32}$ \\[3pt]
%& Hi-counts & 0.00 & $6.66^{+0.01}_{-0.02}$ & $16.19^{+0.36}_{-3.10}$ & $2.51^{+0.32}_{-0.01}$ & $1.70^{+0.21}_{-0.00}$ & $2.26^{+0.74}_{-2.26}$	& $2.17^{+0.83}_{-0.44}$ & $300.20^{+32.08}_{-41.82}$ & $23.77^{+0.35}_{-1.05}$ \\ [3pt]
& Med-flux & 0.00 & $6.82^{+0.01}_{-0.04}$ & $11.58^{+8.42}_{-1.96}$ & $2.71^{+0.38}_{-0.10}$ & $1.91^{+0.39}_{-0.21}$ & $1.26^{+1.74}_{-1.26}$ & $2.89^{+0.11}_{-0.18}$ & $322.86^{+40.24}_{-83.50}$ & $19.70^{+1.85}_{-1.54}$ \\[3pt]
& Lo-flux & 0.00 & $6.62^{+0.05}_{-0.01}$ & $19.99^{+0.01}_{-10.38}$ & $3.03^{+0.07}_{-0.97}$ & $2.00^{+0.24}_{-0.30}$ & $1.55^{+1.45}_{-1.55}$ & $1.75^{+1.25}_{-0.41}$ & $186.99^{+524.52}_{-86.99}$ & $23.36^{+0.00}_{-7.45}$ \\[3pt]
%& Lo-counts & 9.33 & $6.43^{+0.01}_{-0.28}$ & $3.00^{+11.06}_{-0.00}$ & $3.10^{+0.00}_{-0.11}$ & $1.70^{+0.57}_{-0.00}$ & $3.00^{+0.00}_{-3.00}$	& $1.16^{+1.11}_{-0.16}$ & $600.00^{+306.44}_{-134.61}$ & $77.02^{+43.97}_{-8.52}$ \\[3pt]
& 110890101 &	0.00 & $6.99^{+0.01}_{-0.96}$ & $19.34^{+0.66}_{-16.34}$ & $3.00^{+0.10}_{-1.00}$ & $2.37^{+0.53}_{-0.67}$ & $2.69^{+0.31}_{-2.69}$ & $2.94^{+0.06}_{-1.84}$ & $108.76^{+1292.85}_{-8.76}$ & $47.94^{+44.89}_{-22.53}$ \\[3pt]
& 673580101 &	0.39 & $6.65^{+0.29}_{-0.33}$ & $17.99^{+2.01}_{-14.99}$ & $3.10^{+0.00}_{-0.65}$ & $1.70^{+0.90}_{-0.00}$ & $0.00^{+3.00}_{-0.00}$ & $2.22^{+0.78}_{-1.22}$ & $400.00^{+599.99}_{-300}$ & $48.16^{+91.81}_{-33.39}$ \\[3pt]
& 673580201 &	0.21 & $6.41^{+0.01}_{-0.00}$ & $16.00^{+1.43}_{-4.65}$ & $2.55^{+0.45}_{-0.08}$ & $1.73^{+0.42}_{-0.03}$ & $0.00^{+3.00}_{-0.00}$ & $2.82^{+0.18}_{-0.49}$ & $415.47^{+136.63}_{-55.45}$ & $23.06^{+1.99}_{-5.42}$ \\[3pt]	
& 673580301 &	0.07 & $6.51^{+0.33}_{-0.15}$ & $10.36^{+9.64}_{-1.15}$ & $2.56^{+0.54}_{-0.42}$ & $2.12^{+0.53}_{-0.42}$ & $3.00^{+0.00}_{-3.00}$ & $2.98^{+0.02}_{-1.54}$ & $319.56^{+1106.79}_{-95.52}$ & $21.05^{+3.47}_{-4.21}$ \\[3pt]	
& 673580401 &	0.26 & $6.67^{+0.00}_{-0.00}$ & $14.73^{+0.09}_{-0.10}$ & $2.59^{+0.02}_{-0.02}$ & $2.12^{+0.02}_{-0.02}$ & $1.90^{+0.38}_{-0.35}$ & $2.96^{+0.04}_{-0.07}$ & $333.93^{+13.87}_{-14.22}$ & $23.85^{+0.10}_{-0.09}$ \\[3pt]	
& 780560101 &	0.00 & $6.89^{+0.01}_{-0.57}$ & $12.99^{+7.01}_{-9.99}$ & $2.59^{+0.51}_{-0.35}$ & $1.71^{+0.88}_{-0.01}$ & $2.00^{+1.00}_{-2.00}$ & $2.61^{+0.39}_{-1.44}$ & $211.15^{+851.65}_{-91.67}$ & $22.14^{+6.67}_{-3.58}$ \\[3pt]	
& 780561301 &	0.00 & $6.66^{+0.01}_{-0.13}$ & $16.09^{+3.91}_{-4.20}$ & $2.56^{+0.54}_{-0.20}$ & $1.81^{+0.21}_{-0.11}$ & $2.62^{+0.38}_{-2.62}$ & $2.78^{+0.22}_{-0.54}$ & $408.26^{+129.66}_{-91.65}$ & $23.56^{+2.18}_{-5.39}$ \\[3pt]
& 780561401 &	4.71 & $6.38^{+0.16}_{-0.88}$ & $19.00^{+1.00}_{-16.00}$ & $2.20^{+0.90}_{-0.20}$ & $1.70^{+1.20}_{-0.00}$ & $3.00^{+0.00}_{-3.00}$ & $3.00^{+0.00}_{-2.00}$ & $700.00^{+800.00}_{-600.00}$ & $23.17^{+50.09}_{-13.17}$ \\[3pt]	
& 780561501 &	0.00 & $6.80^{+0.03}_{-1.30}$ & $19.50^{+0.50}_{-7.53}$ & $2.54^{+0.56}_{-0.18}$ & $1.96^{+0.11}_{-0.26}$ & $2.20^{+0.28}_{-2.72}$ & $2.82^{+0.18}_{-1.26}$ & $378.09^{+1121.91}_{-224.71}$ & $26.22^{+3.80}_{-8.11}$ \\[3pt]	
& 780561601 &	0.23 & $6.07^{+0.17}_{-0.00}$ & $13.52^{+0.05}_{-0.00}$ & $3.04^{+0.00}_{-0.01}$ & $2.54^{+0.15}_{-0.00}$ & $0.00^{+3.00}_{-0.00}$ & $3.00^{+0.00}_{-0.02}$ & $172.79^{+14.18}_{-45.39}$ & $21.87^{+0.00}_{-0.57}$ \\[3pt]
& 780561701 &	1.55 & $6.73^{+0.16}_{-0.28}$ & $4.96^{+8.33}_{-1.96}$ & $3.10^{+0.00}_{-0.89}$ & $2.00^{+0.72}_{-0.30}$ & $3.00^{+0.00}_{-3.00}$ & $3.00^{+0.00}_{	1.61}$ & $1200.00^{+300.00}_{-299.89}$ & $150.00^{+0.00}_{-82.01}$ \\[3pt]	
& 792180101 & 0.02 & $6.91^{+0.04}_{-0.00}$ & $4.08^{+0.16}_{-1.08}$ & $2.56^{+0.53}_{-0.17}$ & $1.90^{+0.30}_{-0.20}$ & $0.00^{+3.00}_{-0.00}$ & $2.78^{+0.22}_{-0.73}$ & $282.14^{+87.01}_{-136.2}$ & $15.98^{+11.08}_{-0.01}$ \\[3pt]	
& 792180201 &	0.00 & $6.81^{+0.09}_{-0.05}$ & $9.43^{+5.38}_{-0.08}$ & $2.94^{+0.15}_{-0.75}$ & $2.44^{+0.00}_{-0.74}$ & $2.95^{+0.05}_{-2.95}$ & $3.00^{+0.00}_{-1.60}$ & $194.99^{+46.81}_{-29.55}$ & $21.18^{+3.46}_{-1.08}$ \\[3pt]	
& 792180301 &	0.00 & $6.36^{+0.16}_{-0.13}$ & $5.88^{+8.75}_{-2.88}$ & $3.10^{+0.00}_{-0.27}$ & $2.15^{+0.72}_{-0.45}$ & $0.43^{+2.57}_{-0.43}$ & $1.77^{+1.23}_{-0.77}$ & $500.00^{+599.85}_{-300}$ & $106.17^{+33.79}_{-36.99}$ \\[3pt]	
& 792180401 &	5.29 & $6.36^{+0.04}_{-0.00}$ & $4.81^{+1.63}_{-1.81}$ & $3.10^{+0.00}_{-0.72}$ & $1.70^{+0.80}_{-0.00}$ & $3.00^{+0.00}_{-3.00}$ & $2.46^{+0.54}_{-1.03}$ & $958.10^{+541.90}_{-57.75}$ & $150.00^{+0.00}_{-31.88}$ \\[3pt]	
& 792180501 &	2.15 & $6.36^{+0.02}_{-0.01}$ & $5.76^{+1.08}_{-1.17}$ & $3.10^{+0.00}_{-0.52}$ & $1.79^{+0.14}_{-0.09}$ & $3.00^{+0.00}_{-3.00}$ & $3.00^{+0.00}_{-0.88}$ & $902.48^{+0.00}_{-0.00}$ & $150.00^{+0.00}_{-1.59}$ \\[3pt]	
& 792180601 &	1.23 & $6.37^{+0.00}_{-0.06}$ & $4.67^{+2.04}_{-1.67}$ & $3.10^{+0.00}_{-0.81}$ & $2.17^{+0.49}_{-0.47}$ & $3.00^{+0.00}_{-3.00}$ & $3.00^{+0.00}_{-1.29}$ & $1200.00^{+300.00}_{-202.48}$ & $150.00^{+0.00}_{-37.16}$ \\[3pt]	
\hline
\end{tabular}
%\end{adjustbox}
\end{table}

\begin{table}
\centering
\caption[ECM simultaneous results]{The simultaneous fitting results for 1H 0707-495 and IRAS 13224-3809 where the black hole mass $\log (M/M_\odot) = 6.04\pm0.01$  and $6.30\pm0.01$ respectively (to 90\% confidence). Note that zero errors are a result of rounding to 3 sf and also hitting upper or lower model parameter space limits.} \label{SIM-fits}
%\begin{adjustbox}
\begin{tabular}{ccccccccrr}
\hline \\[1pt] 
{Source} & {Obs Id} & {$\chi^2$}  &  {$h_2 (r_g)$} & {$\Gamma_1$}  & {$\Gamma_2$} & {$\log \xi (\text{erg cm s}^{-1})$} & {$b$} & {$t_\text{max} (t_g)$} & {$t_\text{shift} (t_g)$}\\[1pt] 
\hline
\\[2pt] 
1H 0707-495 & 0110890201 & 0.26 & $11.10^{+2.02}_{-0.59}$ & $2.16^{+0.06}_{-0.05}$ & $1.67^{+0.04}_{-0.03}$ & $0.05^{+2.68}_{-0.05}$ & $2.18^{+0.14}_{-0.14}$ & $450.00^{+43.95}_{-73.53}$ & $30.53^{+0.99}_{-3.71}$ \\[3pt] 
& 0148010301 & 0.66 & $7.38^{+0.30}_{-0.27}$ & $2.60^{+0.25}_{-0.45}$ & $1.46^{+0.87}_{-0.06}$ & $0.38^{+0.30}_{-0.27}$ & $1.00^{+.62}_{-0.00}$ & $450.00^{+170.87}_{-81.06}$ & $106.09^{+1.88}_{-1.72}$ \\[3pt] 
& 0506200201 & 0.34 & $11.00^{+1.46}_{-0.35}$ & $2.27^{+0.03}_{-0.00}$ & $1.70^{+0.05}_{-0.03}$ & $2.00^{+0.33}_{-2.00}$ & $3.00^{+0.00}_{-0.14}$ & $488.13^{+30.99}_{-33.09}$ & $22.13^{+0.77}_{-0.78}$ \\[3pt] 
& 0506200301 & 0.49 & $3.00^{+3.48}_{-0.00}$ & $2.00^{+0.35}_{-0.04}$ & $1.40^{+0.03}_{-0.00}$ & $2.41^{+0.17}_{-0.17}$ & $2.75^{+0.25}_{-1.02}$ & $600.05^{+20.62}_{-150.53}$ & $66.77^{+10.94}_{-10.92}$ \\[3pt] 
& 0506200401 & 0.76 & $11.00^{+0.70}_{-0.39}$ & $ 2.55^{+0.02}_{-0.03}$ & $2.16^{+0.02}_{-0.02}$ & $0.95^{+1.74}_{-0.95}$ & $3.00^{+0.00}_{-0.06}$ & $540.99^{+18.15}_{-33.66}$ & $27.91^{+0.84}_{-0.85}$ \\[3pt] 
& 0506200501 & 2.98 & $12.00^{+4.76}_{-9.00}$ & $2.30^{+0.07}_{-0.14}$ & $1.70^{+0.16}_{-0.30}$ & $0.76^{+2.05}_{-0.76}$ & $3.00^{+0.00}_{-0.32}$ & $717.35^{+82.65}_{-99.54}$ & $25.32^{+2.32}_{-4.24}$ \\[3pt] 
& 0511580101 & 0.21 & $11.30^{+0.16}_{-0.13}$ & $2.32^{+0.06}_{-0.04}$ & $1.70^{+0.06}_{-0.03}$ & $0.92^{+1.70}_{-0.92}$ & $3.00^{+0.00}_{-0.18}$ & $643.52^{+66.24}_{-14.28}$ & $21.93^{+1.17 }_{-1.16}$ \\[3pt] 
& 0511580201 & 0.19 & $3.76^{+0.48}_{-0.76}$ & $1.80^{+0.01}_{-0.01}$ & $1.40^{+0.02}_{-0.00}$ & $3.00^{+0.00}_{-0.37}$ & $2.14^{+0.06}_{-0.06}$ & $326.80^{+14.32}_{-14.20}$ & $25.78^{+0.62}_{-0.62}$ \\[3pt] 
& 0511580301 & 0.86 & $3.00^{+0.91}_{-0.00}$ & $1.90^{+0.02}_{-0.01}$ & $1.50^{+0.02}_{-0.03}$ & $3.00^{+0.00}_{-0.24}$ & $3.00^{+0.00}_{-0.15}$ & $501.10^{+63.15}_{-17.67}$ & $21.92^{+0.60}_{-0.51}$ \\[3pt] 
& 0511580401 & 0.95 & $11.45^{+1.08}_{-0.62}$ & $2.48^{+0.02}_{-0.05}$ & $1.71^{+0.06}_{-0.03}$ & $2.36^{+0.20}_{-2.36}$ & $3.00^{+0.00}_{-0.15}$ & $306.33^{+26.94}_{-16.34}$ & $16.73^{+0.54}_{-0.61}$ \\[3pt] 
& 0554710801 & 0.78 & $7.22^{+1.28}_{-1.27}$ & $2.60^{+0.06}_{-0.08}$ & $1.73^{+0.13}_{-0.10}$ & $2.25^{+0.35}_{-2.25}$ & $2.97^{+0.03}_{-0.51}$ & $193.06^{+22.95}_{-23.19}$ & $14.18^{+0.93}_{-1.12}$ \\[3pt] 
& 0653510301 & 0.71 & $20.00^{+0.00}_{-5.87}$ & $2.05^{+0.72}_{-0.33}$ & $1.56^{+1.33}_{-0.14}$ & $0.04^{+2.96}_{-0.04}$ & $3.00^{+0.00}_{-2.00}$ & $193.06^{+233.54}_{-123.73}$ & $81.26^{+32.70}_{-15.10}$ \\[3pt] 
& 0653510401 & 2.07 & $20.00^{+0.00}_{-0.22}$ & $3.30^{+0.00}_{-0.93}$ & $1.51^{+1.11}_{-0.03}$ & $2.25^{+0.75}_{-2.25}$ & $1.00^{+0.32}_{-0.00}$ & $697.70^{+13.26}_{-289.19}$ & $103.03^{+16.97}_{-37.43}$ \\[3pt] 
& 0653510501 & 2.46 & $20.00^{+0.00}_{-2.64}$ & $3.30^{+0.00}_{-1.55}$ & $1.58^{+1.62}_{-0.14}$ & $0.07^{+2.93}_{-0.07}$ & $1.25^{+1.57}_{-0.2}$ & $516.88^{+213.24}_{-341.43}$ & $73.21^{+46.79}_{-73.21}$ \\[3pt] 
& 0653510601 & 2.62 & $20.00^{+0.00}_{-0.39}$ & $3.30^{+0.00}_{-0.73}$ & $1.49^{+0.67}_{-0.03}$ & $0.56^{+1.81}_{-0.56}$ & $2.50^{+0.50}_{-0.98}$ & $505.44^{+199.63}_{-143.04}$ & $94.09^{+9.84}_{-7.24}$ \\[3pt]  
\\[2pt] 
IRAS 13224-3809 & 110890101 & 0.59 &  $9.35^{+10.65}_{-6.35}$ & $2.99^{+0.31}_{-1.19}$ & $1.41^{+1.79}_{-0.01}$ & $3.00^{+0.00}_{-3.00}$ & $3.00^{+0.00}_{-2.00}$ & $190.67^{+609.33}_{-140.67}$ & $45.01^{+44.98}_{-35.01}$ \\[3pt]	 
& 673580101 & 0.43 &  $20.00^{+0.00}_{-17.00}$ & $3.11^{+0.09}_{-0.55}$ & $1.40^{+0.54}_{-0.00}$ & $2.24^{+0.76}_{-1.77}$ & $3.00^{+0.00}_{-0.96}$ & $605.19^{+194.81}_{-70.55}$ & $120.00^{+3.00}_{-7.48}$ \\[3pt]	 
& 673580201 & 0.66 &  $6.11^{+0.40}_{-0.96 }$ & $2.99^{+ 0.15}_{-0.51}$ & $1.40^{+0.54}_{-0.00}$ & $ 3.00^{+0.00}_{-1.57}$ & $2.00^{+0.46}_{-0.49 }$ & $608.45^{+191.55}_{-23.95}$ & $10.00^{+2.75}_{-0.00 }$ \\[3pt]	
& 673580301  & 2.75 &  $14.64^{+5.36}_{-5.32}$ & $1.80^{+0.61}_{-0.00}$ & $3.01^{+0.03}_{-0.71}$ & $0.97^{+2.03}_{-0.97}$ & $3.00^{+0.00}_{-0.61 }$ & $ 109.20^{+70.07}_{-23.74}$ & $16.94^{+3.65}_{-3.93}$ \\[3pt]	 
& 673580401  & 2.25 &  $20.00^{+0.00}_{-3.73 }$ & $1.95^{+0.19}_{-0.15}$ & $3.00^{+0.03}_{-0.17}$ & $0.02^{+2.98}_{-0.02}$ & $1.00^{+0.70 }_{-0.00 }$ & $378.88^{+271.26}_{-224.22}$ & $10.00^{+1.56}_{-0.00}$ \\[3pt]	 
& 780560101 & 0.17 &  $20.00^{+0.00}_{-6.56}$ & $3.16^{+0.04}_{-1.07}$ & $2.19^{+0.14}_{-0.75}$ & $3.00^{+0.00 }_{-3.00}$ & $3.00^{+0.00 }_{-1.63}$ & $663.52^{+111.68}_{-248.94}$ & $20.00^{+5.64}_{-2.96}$ \\[3pt]	 
& 780561301 & 1.97 &  $8.67^{+0.71}_{-0.40 }$ & $3.10^{+0.02}_{-0.39}$ & $1.40^{+0.42}_{-0.00}$ & $3.00^{+0.00}_{-0.78}$ & $1.50^{+0.54 }_{-0.24 }$ & $799.99^{+0.01}_{-183.61}$ & $21.25^{+2.23}_{-4.90}$ \\[3pt]	 
& 780561401 & 0.96 &  $10.71^{+0.82}_{-3.05}$ & $1.80^{+0.10}_{-0.00}$ & $1.89^{+0.11}_{-0.09}$ & $2.23^{+0.67}_{-2.23}$ & $3.00^{+0.00 }_{-0.59}$ & $277.12^{+166.38}_{-77.10}$ & $98.89^{+9.40}_{-15.12}$ \\[3pt]	 
& 780561501 & 0.51 &  $5.12^{+1.76}_{-2.12}$ & $3.10^{+0.03}_{-1.23}$ & $1.40^{+1.42}_{-0.00}$ & $0.73^{+2.27}_{-0.73}$ & $2.50^{+ 0.50 }_{-1.50}$ & $792.89^{+7.11}_{-422.35}$ & $120.74^{+29.26}_{-91.86}$ \\[3pt]	 
& 780561601 & 1.39 &  $7.25^{+1.53}_{-3.91}$ & $3.05^{+0.07}_{-0.46}$ & $2.98^{+0.05}_{-0.28}$ & $0.75^{+2.25}_{-0.75}$ & $3.00^{+0.00}_{-0.50}$ & $200.02^{+60.45}_{-0.02}$ & $91.49^{+2.97}_{-7.94}$ \\[3pt]	 
& 780561701 & 0.50 &  $7.25^{+2.35}_{-2.49}$ & $3.10^{+0.04}_{-1.07}$ & $1.40^{+1.27}_{-0.00}$ & $0.73^{+2.27}_{-0.73}$ & $2.94^{+0.06 }_{-1.94}$ & $717.16^{+82.84}_{-353.04}$ & $120.74^{+29.26}_{-28.95}$ \\[3pt]	 
& 792180101 & 2.47 &  $8.98^{+0.90}_{-1.37}$ & $3.10^{+0.03}_{-0.39}$ & $1.40^{+0.53}_{-0.00}$ & $3.00^{+ 0.00}_{-0.73}$ & $1.75^{+0.99 }_{-0.64}$ & $799.98^{+0.02}_{-144.78}$ & $150.00^{+0.00}_{-44.97}$ \\[3pt]	 
& 792180201 & 2.59 &  $5.99^{+2.64}_{-2.99 }$ & $3.10^{+0.03}_{-0.67}$ & $1.40^{+0.74}_{-0.00}$ & $3.00^{+0.00}_{-3.00}$ & $3.00^{+0.00}_{-0.63}$ & $500.33^{+92.64}_{-33.75}$ & $150.00^{+0.00}_{-29.48}$ \\[3pt]	 
& 792180301 & 0.24 &  $11.40^{+1.29}_{-0.95}$ & $1.88^{+0.41}_{-0.08}$ & $1.40^{+0.25}_{-0.00}$ & $3.00^{+0.00}_{-3.00}$ & $3.00^{+0.00}_{-0.61}$ & $500.33^{+147.38}_{-56.91}$ & $150.00^{+0.00}_{-11.45}$ \\[3pt]	 
& 792180401 & 1.45 &  $7.05^{+0.77}_{-4.05}$ & $3.14^{+0.01}_{-0.02}$ & $1.40^{+1.07}_{-0.00}$ & $3.00^{+0.00}_{-3.00}$ & $2.00^{+1.00 }_{-0.91}$ & $725.08^{+74.92}_{-187.87}$ & $150.00^{+0.00}_{-38.94}$ \\[3pt]	 
& 792180501 & 0.99 &  $8.02^{+0.93}_{-1.13}$ & $3.14^{+0.02}_{-0.01}$ & $1.40^{+0.59}_{-0.00}$ & $3.00^{+0.00}_{-3.00}$ & $2.00^{+0.59 }_{-0.83}$ & $799.85^{+0.15}_{-195.21}$ & $107.18^{+10.64}_{-13.02}$ \\[3pt]	 
& 792180601 & 1.05 &  $8.02^{+0.50}_{-1.62}$ & $3.14^{+0.02}_{-0.01}$ & $1.40^{+1.02}_{-0.00}$ & $3.00^{+0.00}_{-1.00 }$ & $2.00^{+0.58 }_{-0.97}$ & $799.85^{+0.15}_{-373.13}$ & $107.18^{+17.25}_{-9.06}$ \\[3pt]
\hline
\end{tabular}
%\end{adjustbox}
\end{table}

\bsp
\label{lastpage}
\end{document}